\documentclass[journal,comsoc]{IEEEtran}
\usepackage[T1]{fontenc}
\usepackage{amsmath}
\usepackage{amssymb}
\usepackage[cmintegrals]{newtxmath}
\usepackage{booktabs}
\usepackage{graphicx}
\usepackage{subfigure}
\usepackage{caption}
\usepackage{float}
\usepackage{algorithm}  
\usepackage{algpseudocode}
\usepackage{cite}
\usepackage{bm}
\usepackage{color}

\begin{document}
\title{Deep-Reinforcement Learning Multiple Access for Heterogeneous Wireless Networks}

\author{Yiding Yu, Taotao Wang, Soung Chang Liew\\
\thanks{Y. Yu, T. Wang and S. C. Liew are with the Department of Information Engineering, The Chinese University of Hong Kong. Emails:\{yy016, ttwang, soung\}@ie.cuhk.edu.hk}
	%\thanks{The work of S. C. Liew and T. Wang was supported in part by the General Research Funds (Project No. 14200417) and AoE grant E-02/08, established under the University Grant Committee of the Hong Kong Special Administrative Region, China.}
}
\maketitle
\pagestyle{empty}
\thispagestyle{empty} 

\begin{abstract}
This paper investigates the use of deep reinforcement learning (DRL) in a MAC protocol for heterogeneous wireless networking referred to as Deep-reinforcement Learning Multiple Access (DLMA). The thrust of this work is partially inspired by the vision of DARPA SC2, a 3-year competition whereby competitors are to come up with a clean-slate design that ``best share spectrum with any network(s), in any environment, without prior knowledge, leveraging on machine-learning technique''.  Specifically, this paper considers the problem of sharing time slots among a multiple of time-slotted networks that adopt different MAC protocols. One of the MAC protocols is DLMA. The other two are TDMA and ALOHA. The nodes operating DLMA do not know that the other two MAC protocols are TDMA and ALOHA. Yet, by a series of observations of the environment, its own actions, and the resulting rewards, a DLMA node can learn an optimal MAC strategy to coexist harmoniously with the TDMA and ALOHA nodes according to a specified objective (e.g., the objective could be the sum throughput of all networks, or a general  $ \alpha $-fairness objective). 
%Beyond the contribution to wireless networking, in the course of doing this work, we discovered a way to generalize the Q-learning framework (Q-learning is a subclass of reinforcement learning) so that it can achieve more general objectives than the conventional Q-learning can.
\end{abstract}

%\begin{IEEEkeywords}
%Communications Society, IEEE, IEEEtran, journal, \LaTeX, paper, template.
%\end{IEEEkeywords}

% \fi{}
\IEEEpeerreviewmaketitle

\section{Introduction}
This paper investigates a new generation of wireless multiple access control (MAC) protocol that leverages the latest advances in ``deep reinforcement learning''. The work is partially inspired by our participation in the Spectrum Collaboration Challenge (SC2), a three-year competition hosted by DARPA of the United States \cite{DARPAwebsite}.\footnote{Two of the authors are currently participating in this competition.} Quoting DARPA, ``SC2 is the first-of-its-kind collaborative machine-learning competition to overcome scarcity in the radio frequency (RF) spectrum. Today, spectrum is managed by dividing it into rigid, exclusively licensed bands. In SC2, competitors will reimagine a new, more efficient wireless paradigm in which radio networks autonomously collaborate to dynamically determine how the spectrum should be used moment to moment.'' In other words, DARPA aims for a clean-slate design in which different wireless networks share spectrum in a very dynamic manner based on instantaneous supply and demand. In DARPA's vision, ``winning design is the one that best share spectrum with any network(s), in any environment, without prior knowledge, leveraging on machine-learning technique''. DARPA's vision necessitates a total re-engineering of the PHY, MAC, and Network layers of wireless networks. 

As a first step, this paper investigates a new MAC design that exploits deep Q-network (DQN) algorithm \cite{DQNpaper}, a deep reinforcement learning (DRL) algorithm that combines deep neural networks \cite{lecun2015deep} with Q-learning \cite{watkins1992q}. DQN was shown to be able to achieve superhuman-level playing performance in video games. Our MAC design aims to learn an optimal way to use the time-spectrum resources by a series of observations and actions without the need to know the operating mechanisms of the MAC protocols of other coexisting networks. In particular, our MAC strives to achieve optimal performance as if it knew the MAC protocols of these networks in detail. In this paper, we refer to our MAC protocol as deep-reinforcement learning multiple access (abbreviated as DLMA), and a radio node operating DLMA as a DRL agent.  

For a focus, this paper considers time-slotted systems and the problem of sharing the time slots among multiple wireless networks. In general, DLMA can adopt different objectives in time-slot sharing. We first consider the objective of maximizing the sum throughput of all the networks. We then reformulate DLMA to achieve a general  $ \alpha $-fairness objective. In particular, we show that DLMA can achieve near-optimal sum throughput and proportional fairness when coexisting with a TDMA network, an ALOHA network, and a mix of TDMA and ALOHA networks, without knowing the coexisting networks are TDMA and ALOHA networks. Learning from the experience it gathers from a series of state-action-reward observations, a DRL agent tunes the weights of the neural network within its MAC machine to zoom to an optimal MAC strategy.

This paper also addresses the issue of why DRL is preferable to the traditional reinforcement learning (RL) \cite{sutton1998reinforcement} for wireless networking. Specifically, we demonstrate that the use of deep neural networks (DNN) in DRL affords us with two essential properties to wireless MAC: (i) fast convergence to near-optimal solutions; (ii) robustness against non-optimal parameter settings (i.e., fine parameter tuning and optimization are unnecessary with our DRL framework). Compared with MAC based on traditional RL, DRL converges faster and is more robust. Fast convergence is critical to wireless networks because the wireless environment may change quickly as new nodes arrive, and existing nodes move or depart. If the environmental ``coherence time'' is much shorter than the convergence time of the wireless MAC, the optimal strategy would elude the wireless MAC as it continuingly tries to catch up with the environment. Robustness against non-optimal parameter settings is essential because the optimal parameter settings for DRL (and RL) in the presence of different coexisting networks may be different. Without the knowledge of the coexisting networks, DRL (and RL) cannot optimize its parameter settings a priori. If non-optimal parameter setting can also achieve roughly the same optimal throughput at roughly the same convergence rate, then optimal parameter settings are not essential for practical deployment.

In our earlier work \cite{yu2018deep}, we adopted a plain DNN as the neural network in our DRL overall framework. In this work, we adopt a deep residual network (ResNet) \cite{he2016deep}. The results of all sections in the current paper are based on ResNet, except Section \ref{sec:ResNet} where we study deep ResNet versus plain DNN.  A key advantage of ResNet over plain DNN is that the same static ResNet architecture can be used in DRL for different wireless  network scenarios; whereas for plain DNN, the optimal neural network depth varies from case to case.

Overall, our main contributions are as follows:
\begin{itemize}
\item 	We employ DRL for the design of DLMA, a MAC protocol for heterogeneous wireless networking. Our DLMA framework is formulated to achieve general  $ \alpha $-fairness among the heterogeneous networks.  Extensive simulation results show that  DLMA can achieve near-optimal sum throughput and proportional fairness objectives. In particular, DLMA achieves these objectives without knowing the operating mechanisms of the MAC protocols of the other coexisting networks. 
\item We demonstrate the advantages of exploiting DRL in heterogeneous wireless networking compared with the traditional RL method.  In particular, we show that DRL can accelerate convergence to an optimal solution and is more robust against non-optimal parameter settings, two essential properties for practical deployment of DLMA in real wireless networks.  
\item In the course of our generalization to the  $ \alpha $-fairness objective in wireless networking, we discovered an approach to generalize the Q-learning framework so that more general objectives can be achieved. In particular, we argue that for generality, we need to separate the Q function and the objective function upon which actions are chosen to optimize -- in conventional Q learning, the Q function itself is the objective function. We give a framework on how to relate the objective function and the Q function in the general set-up. 
\end{itemize}
\subsection{Related Work}
RL is a machine-learning paradigm, where agents learn successful strategies that yield the largest long-term reward from trial-and-error interactions with their environment \cite{sutton1998reinforcement}. The most representative RL algorithm is the Q-learning algorithm \cite{watkins1992q}. Q-learning can learn a good policy by updating an action-value function, referred to as the Q function, without an operating model of the environment. When the state-action space is large and complex, deep neural networks can be used to approximate the Q function and the corresponding algorithm is called DRL \cite{DQNpaper}. This work employs DRL to speed up convergence and increase the robustness of DLMA (see our results Section \ref{RLvsDRL}).

RL was employed to develop channel access schemes for cognitive radios \cite{li2010multiagent,yau2010enhancing,wu2010spectrum,bkassiny2011distributed} and wireless sensor networks \cite{liu2006rl,chu2012aloha}. Unlike this paper, these works do not leverage the recent advances in DRL.

There has been little prior work exploring the use of DRL to solve MAC problems, given that DRL itself is a new research topic. The MAC scheme in \cite{naparstek2017deep} employs DRL in homogeneous wireless networks. Specifically, \cite{naparstek2017deep} considered a network in which $N$  radio nodes dynamically access  $K$ orthogonal channels using the same DRL MAC protocol. By contrast, we are interested in heterogeneous networks in which the DRL nodes must learn to collaborate with nodes employing other MAC protocols.

The authors of \cite{wang2018deep} proposed a DRL-based channel access scheme for wireless sensor networks. Multiple frequency channels were considered. In RL terminology, the multiple frequency channels with the associated Markov interference models form the ``environment'' with which the DRL agent interacts. There are some notable differences between \cite{wang2018deep} and our investigation here. The Markov environmental model in \cite{wang2018deep} cannot capture the interactions among nodes due to their MAC protocols. In particular, the Markov environmental model in \cite{wang2018deep} is a ``passive'' model not affected by the ``actions'' of the DRL agent. For example, if there is one exponential backoff ALOHA node (see Section \ref{system_model} for definition) transmitting on a channel, the collisions caused by transmissions by the DRL agent will cause the channel state to evolve in intricate ways not captured by the model in \cite{wang2018deep}.

In \cite{challita2017deep}, the authors employed DRL for channel selection and channel access in LTE-U networks. Although it also aims for heterogeneous networking in which LTE-U base stations coexist with WiFi APs, its focus is on matching downlink channels to base stations; we focus on sharing an uplink channel among users. More importantly, the scheme in \cite{challita2017deep} is model-aware in that the LTE-U base stations know that the other networks are WiFi. For example, it uses an analytical equation (equation (1) in \cite{challita2017deep}) to predict the transmission probability of WiFi stations. By contrast, our DLMA protocol is model-free in that it does not presume knowledge of coexisting networks and is outcome-based in that it derives information by observing its interactions with the other stations in the heterogeneous environment.

\section{DLMA Protocol}\label{DLMA_protocol}
This section first introduces the time-slotted heterogeneous wireless networks considered in this paper. Then a short overview of RL is given. After that, we present our DLMA protocol, focusing on the objective of maximizing the sum throughput of the overall system. A generalized DLMA protocol that can achieve  $ \alpha $-fairness objective will be given in Section \ref{general_objective}. 
\subsection{Time-Slotted Heterogeneous Wireless Networks}\label{system_model}
We consider time-slotted heterogeneous wireless networks in which different radio nodes transmit packets to an access point (AP) via a shared wireless channel, as illustrated in Fig. \ref{fig:system_model}. We assume all the nodes can begin transmission only at the beginning of a time slot and must finish transmission within that time slot. Simultaneous transmissions of multiple nodes in the same time slot result in collisions. The nodes may not use the same MAC protocol: some may use TDMA and/or ALOHA, and at least one node uses our proposed DLMA protocol. The detailed descriptions of different radio nodes are given below:
\begin{figure}[!t]
	\centering
	\includegraphics[scale=0.03]{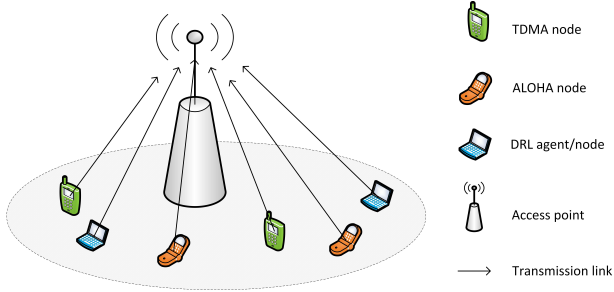}
	\caption{A heterogeneous multiple-access system with a mix of DRL nodes and other nodes.}
	\label{fig:system_model}
	%\vspace{-0.22in}
\end{figure}
\begin{itemize}
\item \textbf{TDMA:} The TDMA node transmits in  $ X $ specific slots within each frame of $ Y $ slots in a repetitive manner from frame to frame.
\item \textbf{$  q$-ALOHA:}  A  $ q $-ALOHA node transmits with a fixed probability  $ q $ in each time slot in an i.i.d. manner from slot to slot. 
\item \textbf{Fixed-window ALOHA:} A fixed-window ALOHA (FW-ALOHA) node generates a random counter value $ w $  in the range of  $ \left[0, W-1 \right]  $  after it transmits in a time slot. It then waits for  $ w $ slots before its next transmission. The parameter $ W $  is referred to as the window size. 
\item \textbf{Exponential backoff ALOHA:} Exponential backoff ALOHA (EB-ALOHA) is a variation of window-based ALOHA in which the window size is not fixed. Specifically, an EB-ALOHA node doubles its window size each time when its transmission encounters a collision, until a maximum window size $ 2^mW $  is reached, where  $ m $ is the ``maximum backoff stage''. Upon a successful transmission, the window size reverts back to the initial window size  $ W $.
\item \textbf{DRL agent/node:} A DRL agent/node is the radio node that adopts our DLMA protocol. For a DRL node, if it transmits, it will get an immediate ACK from the AP, indicating whether the transmission is successful or not; if it does not transmit, it will listen to the channel and get an observation from the environment, indicating other nodes' transmission results or idleness of the channel. Based on the observed results, the DRL node can set different objectives, such as maximizing the sum throughput of the overall system (as formulated in Part C of this section) and achieving a general  $ \alpha $-fairness objective (as formulated in Section \ref{general_objective}).
\end{itemize}

\subsection{Overview of RL}\label{RLoverview}
\begin{figure}[!t]
	\centering
	\includegraphics[scale=0.05]{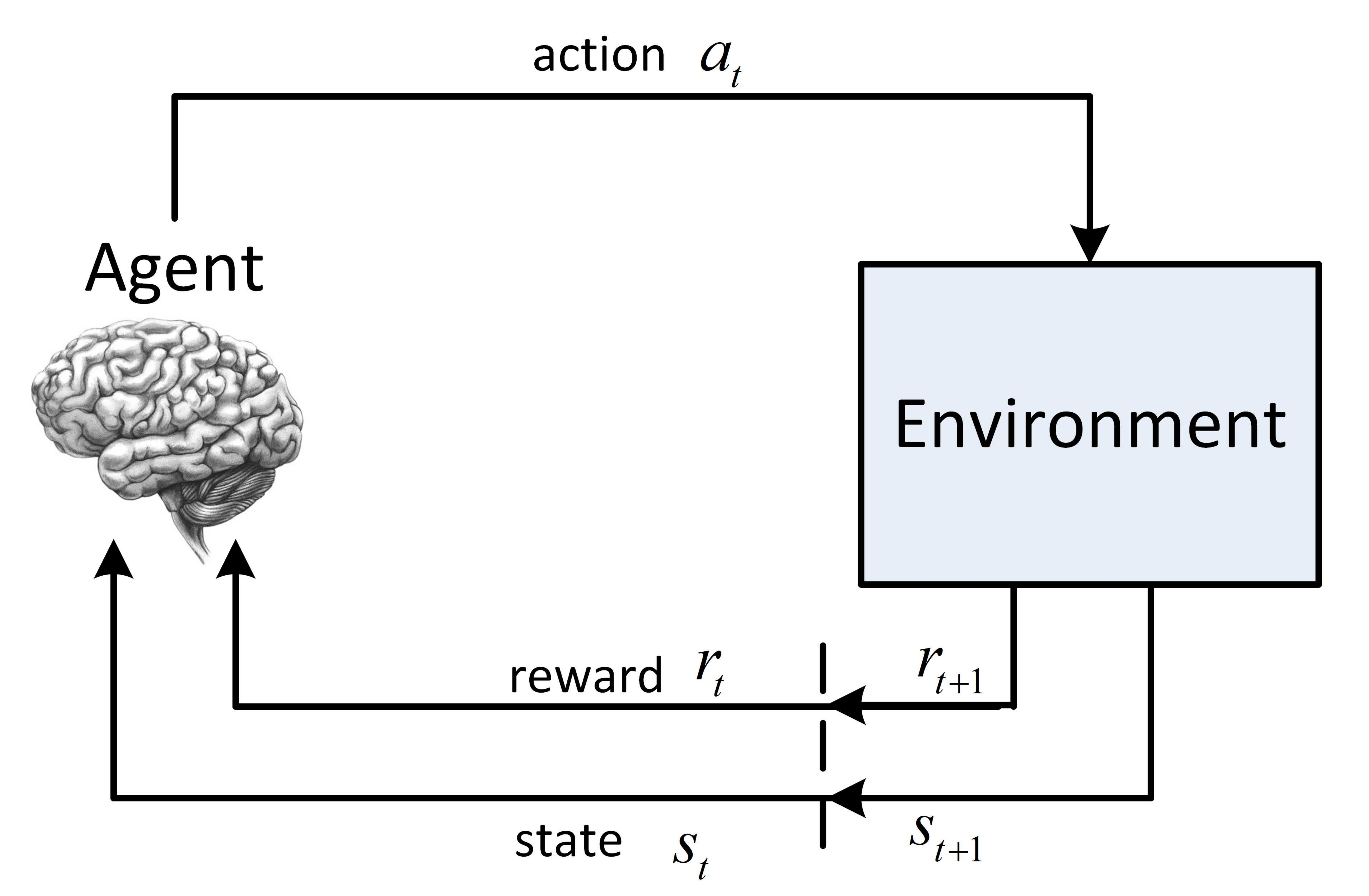}
	\caption{The agent-environment interaction process.}
	\label{fig:RL}
	%\vspace{-0.22in}
\end{figure}

In RL \cite{sutton1998reinforcement}, an agent interacts with an environment in a sequence of discrete times,  $ t=0,1,2, \cdots $, to accomplish a task, as shown in Fig. \ref{fig:RL}. At time $ t $, the agent observes the state of the environment  $ s_t \in S $, where $ S $ is the set of possible states. It then takes an action $ a_t \in A_{s_t} $,  where  $  A_{s_t} $ is the set of possible actions at state $ s_t $. As a result of the state-action pair $ \left( s_t, a_t\right)  $, the agent receives a reward $ r_{t+1} $, and the environment moves to a new state $ s_{t+1} $ at time  $ t+1 $. The goal of the agent is to effect a series of rewards  $ \left\lbrace r_t \right\rbrace_{t=1,2, ...}  $  through its actions to maximize some performance criteria. For example, the performance criterion to be maximized at time $ t $ could be $ {R_t} \buildrel \Delta \over = \sum\nolimits_{\tau  = t}^\infty  {{\gamma ^{\tau  - t}}{r_{\tau {\rm{ + }}1}}}  $, where $ \gamma \in \left( 0, 1\right]  $  is a discount factor for weighting future rewards. In general, the agent takes actions according to some decision policy  $ \pi $. RL methods specify how the agent changes its policy as a result of its experiences. With sufficient experiences, the agent can learn an optimal decision policy $ \pi^* $ to maximize the long-term accumulated reward\cite{sutton1998reinforcement}. 

Q-learning \cite{watkins1992q} is one of the most popular RL methods. A Q-learning RL agent learns an action-value function ${Q^\pi }(s,a)$ corresponding to the expected accumulated reward when an action $ a $ is taken in the environmental state $ s $ under the decision policy $ \pi $:
\begin{equation}
{Q^\pi }\left( {s,a} \right) \buildrel \Delta \over = \bm{E}\left[ {{R_t}\left| {{s_t} = s,{a_t} = a,\pi } \right.} \right].
\end{equation}
The optimal action-value function, $ {Q^ * }\left( {s,a} \right) \buildrel \Delta \over = {\max _\pi }{Q^\pi }\left( {s,a} \right) $, obeys the Bellman optimality equation\cite{sutton1998reinforcement}:
\begin{equation}
{Q^ * }\left( {s,a} \right) = {\bm{E}_{s'}}\left[ {{r_{t+1}} + \gamma \mathop {\max }\limits_{a'} {Q^*}\left( {s',a'} \right)| s_t=s ,a_t=a} \right],
\end{equation}
where $ s' $ is the new state after the state-action pair $ \left( s, a\right)  $. The main idea behind Q-learning is that we can iteratively estimate $ {Q^ * }\left( {s,a} \right) $ at the occurrences of each state-action pair $ \left( s, a\right)  $. 

Let $ q\left( s, a\right)  $ be the estimated action-value function during the iterative process. Upon a state-action pair $({s_t},{a_t})$ and a resulting reward $ r_{t+1} $, Q-learning updates $q({s_t},{a_t})$ as follows:
\begin{align}\label{Qupdate}
q\left( {{s_t},{a_t}} \right) &\leftarrow q\left( {{s_t},{a_t}} \right) +\nonumber \\
&\beta \left[ {{r_{t + 1}} + \gamma \mathop {\max }\limits_{a'} q\left( {{s_{t + 1}},a'} \right) - q\left( {{s_t},{a_t}} \right)} \right],
\end{align}
where $ \beta  \in (0,1] $ is the learning rate.

While the system is updating $ q\left( {s,a} \right) $, it also makes decisions based on  $ q\left( {s,a} \right) $. The $ \varepsilon $-greedy policy is often adopted, i.e.,  
%For the  $ \varepsilon $-greedy policy, the agent selects the greedy action $ {a_t}=\arg {\max _a}q\left( {{s_t},a} \right) $ with probability  $ 1-\varepsilon $, and selects a random action with probability $ \varepsilon $.
\begin{equation}
{a} = \left\{ {\begin{array}{*{20}{c}}
	{\mathop {\arg \max_{\tilde a} } q\left( {{s},{\tilde a}} \right),}&{\text{with probability} \ 1-\varepsilon},\\
	{\text{random action},}&{\text{witih probability} \ \varepsilon}.
	\end{array}} \right.
\end{equation}
A reason for randomly selecting an action is to avoid getting stuck with a $ q\left( {s,a} \right) $ function that has not yet converged to $ {Q^ * }\left( {s,a} \right) $. 
\subsection{DLMA Protocol Using DRL}\label{DLMA_formulation}
This subsection describes the construction of our DLMA protocol using the DRL framework. 

The \textit{action} taken by a DRL agent at time $ t $ is $ {a_t} \in$  \{\textit{TRANSMIT, WAIT}\}, where \textit{TRANSMIT} means that the agent transmits, and \textit{WAIT} means that the agent does not transmit. We denote the \textit{channel observation} after taking action $ a_t $ by $ {z_t} \in$ \{\textit{SUCCESS}, \textit{COLLISION}, \textit{IDLENESS}\}, where \textit{SUCCESS} means one and only one station transmits on the channel; \textit{COLLISION} means multiple stations transmit, causing a collision; \textit{IDLENESS} means no station transmits. The DRL agent determines $ z_t $  from an ACK signal from the AP (if it transmits) and listening to the channel (if it waits). We define the \textit{channel state} at time $ t + 1 $ as the action-observation pair $ {c_{t+1}} \buildrel \Delta \over = \left( {{a_t},{z_t}} \right) $. There are five possibilities for  $ c_{t+1}$: \{\textit{TRANSMIT, SUCCESS}\}, \{\textit{TRANSMIT, COLLISION}\}, \{\textit{WAIT, SUCCESS}\}, \{\textit{WAIT, COLLISION}\} and \{\textit{WAIT, IDLENESS}\}.  We define the \textit{environmental state} at time $  t + 1  $ to be $ {s_{t+1}} \buildrel \Delta \over = \left\{ {{c_{t - M+2 }},...,{c_{t}},{c_{t+1}}} \right\} $, where the parameter $ M $  is the state history length to be tracked by the agent. After taking action $ a_t $, the transition from state $ s_t $ to $ s_{t+1} $ generates a \textit{reward} $ r_{t+1} $, where $ r_{t+1}=1 $ if $ z_t= $ \textit{SUCCESS}; $ r_{t+1}=0 $ if $ z_t=$ \textit{COLLISION} or \textit{IDLENESS}. The definition of reward here corresponds to the objective of maximizing the sum throughput. We define a reward vector in Section \ref{general_objective}  so as to generalize DLMA to achieve the  $ \alpha $-fairness objective.  

So far, the above definitions of ``action'', ``state'' and ``reward'' also apply to an RL agent that adopts (\ref{Qupdate}) as its learning algorithm. We next motivate the use of DRL and then provide the details of its use. 

Intuitively, subject to a non-changing or slow-changing environment, the longer the state history length $ M $, the better the decision can be made by the agent. However, a large $ M $ induces a large state space for the RL algorithm. With a large number of state-action entries to be tracked, the step-by-step and entry-by-entry update $ q\left( s, a\right)  $ is very inefficient. To get a rough idea, suppose that $ M=10 $ (a rather small state history to keep track of), then there are $ 5^{10}\approx 10 $ million possible values for state $ s $. Suppose that for convergence to the optimal solution, each state-action value must be visited at least once. If each time slot is 1 $ ms $ in duration (typical wireless packet transmission time), the convergence of RL will take at least $ 5^{10} \times 2 $ $ ms $, or more than 5 hours. Due to node mobility, arrivals and departures, the wireless environment will most likely to have changed well before then. Section \ref{RLvsDRL} of this paper shows that applying  DRL to DLMA accelerates the convergence speed significantly (convergence is obtained in seconds, not hours).

In DRL, a deep neural network \cite{lecun2015deep} is used to approximate the action-value function,  $q\left( {s,a;{\bm{\theta }}} \right) \approx {Q^*}\left( {s,a} \right)$, where $q\left( {s,a;{\bm{\theta }}} \right)$  is the approximation given by the neural network and $ \bm{\theta } $  is a parameter vector containing the weights of the edges in the neural network. The input to the neural network is a state  $ s $, and the outputs are approximated  $ q $ values for different actions   $ \mathbb{Q}= \left\{ {q\left( {s,a;{\bm{\theta }}} \right)|a \in {A_s}} \right\}$. We refer to the neural network as the Q neural network (QNN) and the corresponding RL algorithm as DRL. Rather than following the tabular update rule of the traditional RL in (\ref{Qupdate}), DRL updates  $q\left( {s,a;{\bm{\theta }}} \right)$  by adjusting the $ \bm{\theta } $  in the QNN through a training process.

In particular, QNN is trained by minimizing prediction errors of  $q\left( {s,a;{\bm{\theta }}} \right)$. Suppose that at time  $ t $, the state is  $ s_t $  and the weights of QNN are $ \bm{\theta } $. The DRL agent takes an action  ${a_t} = \arg {\max _a}q\left( {{s_t},a;{\bm{\theta }}} \right)$, where   $q\left( {s_t,a;{\bm{\theta }}} \right)$  for different actions  $ a $  are given by the outputs of QNN. Suppose that the resulting reward is  $ r_{t+1} $  and the state moves to   $ s_{t+1} $. Then, $\left( {{s_t},{a_t},{r_{t + 1}},{s_{t + 1}}} \right)$  constitutes an ``experience sample'' that will be used to train the QNN.  For training, we define the prediction error of QNN for the particular experience sample  $\left( {{s_t},{a_t},{r_{t + 1}},{s_{t + 1}}} \right)$  to be
\begin{equation}\label{loss1}
{L_{{s_t},{a_t},{r_{t + 1}},{s_{t + 1}}}}\left( {{\bm{\theta}}} \right) = {\left( {y_{{r_{t + 1}},{s_{t + 1}}}^{QNN} - q\left( {{s_t},{a_t};{\bm{\theta}}} \right)} \right)^2},
\end{equation}
where $ \bm{\theta } $ are the weights in QNN, $q\left( {s_t,a_t;{\bm{\theta }}} \right)$  is the approximation given by QNN, and  $y_{{r_{t + 1}},{s_{t + 1}}}^{QNN}$  is the target output for QNN given by 
\begin{equation}\label{yQNN1}
y_{{r_{t + 1}},{s_{t + 1}}}^{QNN} = {r_{t + 1}} + \gamma {\max _{a'}}q\left( {{s_{t + 1}},a';{\bm{\theta }}} \right).
\end{equation}
Note that  $y_{{r_{t + 1}},{s_{t + 1}}}^{QNN}$  is a refined target output based on the current reward $ r_{t+1} $  plus the predicted discounted rewards going forward  $\gamma {\max _{a'}}q\left( {{s_{t + 1}},a';{\bm{\theta }}} \right)$ given by QNN. We can train QNN, i.e., update $ \bm{\theta } $, by applying a semi-gradient algorithm \cite{sutton1998reinforcement} in (\ref{loss1}). The iteration process of $ \bm{\theta } $ is given by
\begin{equation}\label{theta_update1}
\text{Iterate} \ {\bm{\theta}} \leftarrow {\bm{\theta }} - \rho \left[ {y_{{r_{t + 1}},{s_{t + 1}}}^{QNN} - q\left( {{s_t},{a_t};{\bm{\theta}}} \right)} \right]\nabla q\left( {{s_t},{a_t};{\bm{\theta }}} \right),
\end{equation}
where  $\rho $ is the step size in each adjustment.

For algorithm stability,  the ``experience replay'' and ``quasi-static target network'' techniques can be used\cite{DQNpaper}. For ``experience replay'', instead of training QNN with a single experience at the end of each execution step, we could pool together many experiences for batch training. In particular, an experience memory with a fixed storage capacity is used for storing the experiences $e = \left( {s,a,r,s'} \right)$  gathered from different time steps in an FIFO manner, i.e., once the experience memory is full, the oldest experience is removed from, and the new experience is put into, the experience memory. For a round of training, a minibatch  $ E $ consisting of $ N_E $  random experiences are taken from the experience memory for the computation of the loss function.  For ``quasi-static target network'', a separate target QNN with parameter $ \bm{\theta^- } $ is used as the target network for training purpose. Specifically, the   $q\left( \cdot \right)$ in (\ref{yQNN1}) is computed based on this separate target QNN, while the $q\left( \cdot \right)$  in (\ref{loss1}) is based on QNN under training. The target QNN is a copy of an earlier QNN: every  $ F $ time steps, the target QNN is replaced by the latest QNN, i.e., setting $ \bm{\theta^- } $ to the latest  $ \bm{\theta } $  of QNN. With these two techniques, equations (\ref{loss1}), (\ref{yQNN1}), and (\ref{theta_update1}) replaced by the following: 

\begin{equation}\label{loss2}
{L_E}\left( {{\bm{\theta }}} \right) = \frac{1}{{{N_E}}}\sum\limits_{e \in {E}} {{{\left( {y_{r,s'}^{QNN} - q\left( {s,a;{\bm{\theta }}} \right)} \right)}^2}}, 
\end{equation}
\begin{equation}\label{yQNN2}
y_{r,s'}^{QNN} = r + \gamma {\max _{a'}}q\left( {s',a';{\bm{\theta^- }}} \right),
\end{equation}
\begin{align}\label{theta_update2}
\text{Iterate} \ {\bm{\theta }} \leftarrow {\bm{\theta }} - \frac{\rho }{{{N_E}}} \sum\limits_{e \in {E}} {\left[ {y_{r,s'}^{QNN} - q\left( {s,a;{\bm{\theta }}} \right)} \right]\nabla q\left( {s,a;{\bm{\theta }}} \right)}, 
\end{align}
\begin{equation}
\text{Every} \ F \ \text{steps, set}  \ \bm{\theta^- } \ \text{to} \ \bm{\theta }.
\end{equation}
The pseudocode of DLMA algorithm is given in Algorithm \ref{alg:DLMA1}.

\begin{algorithm}
	\caption{DLMA with the sum throughput objective}\label{alg:DLMA1}
	\begin{algorithmic}
		\State Initialize $ s_0 $, $ \varepsilon $, $ \gamma $, $ \rho $,  $ N_E $, $ F $
		\State Initialize experience memory $ EM $
		\State Initialize the parameter of QNN as $  \bm{\theta } $ 
		\State Initialize the parameter of target QNN  $ \bm{\theta^- }=\bm{\theta } $ 
		\For{$ t=0,1,2, \cdots $ in DLMA}
		%\State \textit{// Execution}
		\State Input $ s_t $ to QNN and output $\mathbb{Q} = \left\{ {q\left( {s_t,a,{\bm{\theta}}} \right)|a \in {A_{s_t}}} \right\}$
		\State Generate action $ a_t $ from $ \mathbb{Q} $ using $ \varepsilon $-greedy algorithm
		\State Observe $ z_t $, $ r_{t+1} $ 
		\State Compute $ s_{t+1} $ from $ s_t $, $ a_t $ and $ z_t $
		%\If{$ \vert E \vert <  N_{max}  $}
		%	\State Add $ e =  \left( {s, a, r, s'} \right) $ to  $ E $
		%\Else
		%	\State Add $ e $ to $ E $ in an FIFO manner
		%\EndIf
		\State Store $ \left( {s_t, a_t, r_{t+1}, s_{t+1}} \right) $ to $ EM $
		\State \textbf{if} Remainder($ t/F==0 $) \textbf{then}  $ I=1 $ \textbf{else} $ I=0 $
		\State \Call{TrainQNN}{$ \gamma $, $ \rho $, $ N_E $, $ I $, $ EM $, $ \bm{\theta} $, $ \bm{\theta^-} $}\footnotemark
		%\State \textit{// Training }
		\EndFor \\
		\Procedure{TrainQNN}{$ \gamma $, $ \rho $, $ N_E $, $ I $, $ EM $, $ \bm{\theta} $, $ \bm{\theta^-} $}
		\State Randomly sample $ N_E $ experience tuples from $ EM $ as $ E $
		\For{each sample $ e = \left( {s,a,r,s'} \right) $ in $ E $}
		\State Calculate $ y_{r,s'}^{QNN} = r + \gamma \mathop {\max }\limits_{a'} q\left( {s',a';{\bm{\theta^- }}} \right) $
		\EndFor 
		\State Perform Gradient Descent to update  $ \bm{\theta } $ in QNN:
		\begin{equation*}
		\text{Iterate} \ {\bm{\theta }} \leftarrow {\bm{\theta }} - \frac{\rho }{{{N_E}}}\sum\limits_{e \in E} {\left[ {y_{r,s'}^{QNN} - q\left( {s,a;{\bm{\theta }}} \right)} \right]\nabla q\left( {s,a;{\bm{\theta }}} \right)}
		\end{equation*}
		\If{$ I==1 $}
		\State Update $ \bm{\theta^- } $ in target QNN by setting $ \bm{\theta^-} = \bm{\theta } $
		\EndIf
		\EndProcedure
	\end{algorithmic}
\end{algorithm}
\footnotetext{For convenience, in our simulation, we assume execution of decisions and training of QNN are run synchronously. In particular, the training is done at the end of each time step after an execution. In practice, for efficiency and to allow more time for training, execution and training can be done asynchronously and in parallel. In this case, the  experiences resulted from executions in successive time steps are fed to the experience memory in a continuous manner. Meanwhile, training takes random minibatches of experiences from the experience memory in a parallel and asynchronous fashion continuously. There could be more than one training round (i.e., more than one minibatches used for training ) per execution time step if sufficient computation resources are available. Once in a while, the QNN used in execuction is replaced by the newly trained QNN by substituting the $ \bm{\theta } $ in QNN with the new $ \bm{\theta } $ in the newly trained QNN; at the same time $ \bm{\theta^- } $ in the target QNN is also replaced by the new $ \bm{\theta } $ for future training purposes. }

\section{Sum Throughput Performance Evaluation}\label{exp_sum}
This section investigates the performance of DLMA  with the objective of maximizing the sum throughput of all the coexisting networks. For our investigations, we consider the interactions of DRL nodes with TDMA nodes, ALOHA nodes, and a mix of TDMA nodes and ALOHA nodes. Section \ref{general_objective} will reformulate the DLMA framework to achieve a general $ \alpha $-fairness objective (which also includes maximizing sum-throughput objective as a subcase); Section \ref{exp_pro} will present the corresponding results.

\begin{figure}[!t]
	\centering
	\includegraphics[scale=0.045]{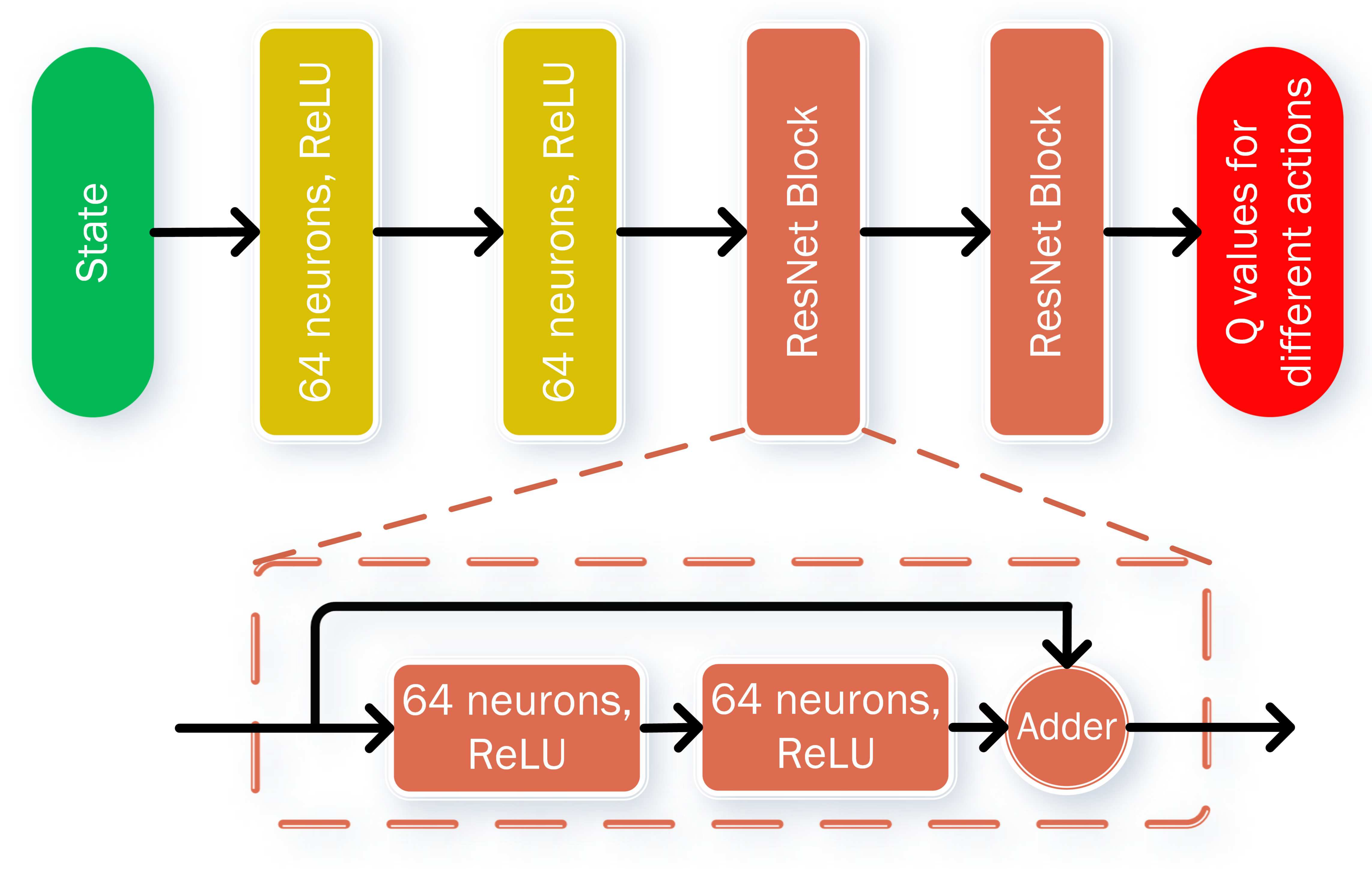}
	\caption{ResNet Architecture}
	\label{ResNetArchitecture}
	%\vspace{-0.22in}
\end{figure}

As illustrated in Fig. \ref{ResNetArchitecture}, the architecture of QNN used in DLMA is a six-hidden-layer ResNet with 64 neurons in each hidden layer. The activation functions used for the neurons are \textit{ReLU} functions \cite{lecun2015deep}. The first two hidden layers of QNN are fully connected, followed by two ResNet blocks. Each ResNet block contains two fully connected hidden layers plus one ``shortcut'' from the input to the output of the ResNet block.  The state, action and reward of DRL follow the definitions in Section \ref{DLMA_formulation}. The state history length  $ M $ is set to 20, unless stated otherwise. When updating the weights  $ \bm{\theta} $ of QNN, a minibatch of 32 experience samples are randomly selected from an  experience-replay reservoir of 500 prior experiences for the computation of the loss function (\ref{loss2}). The experience-replay reservoir is updated in a FIFO manner: a new experience replaces the oldest experience in it.  The \textit{RMSProp} algorithm \cite{tieleman2012lecture} is used to conduct minibatch gradient descent for the update of $\bm{\theta} $.  To avoid getting stuck with a suboptimal decision policy before sufficient learning experiences,  we apply an exponential decay  $ \varepsilon $-greedy algorithm:  $ \varepsilon $ is initially set to 0.1 and decreases at a rate of 0.995 every time slot until its value reaches 0.005. A reason for not decreasing $ \varepsilon $  all the way to zero is that in a general wireless setting, the wireless environment may change dynamically with time (e.g., nodes are leaving and joining the network). Having a positive $ \varepsilon $ at all time allows the decision policy to adapt to future changes. Table \ref{tab:hyperparameters} summarizes the hyper-parameter settings in our investigations. 
\begin{table}[!htbp]
	\centering
	\caption{DLMA Hyper-parameters}\label{tab:hyperparameters}
	\vspace{0in}
	\begin{tabular}{ccc}
		\toprule
		Hyper-parameters& Value\\
		\midrule
		State history length  $ M $& 20, unless stated otherwise\\
		Discount factor $ \gamma $ & 0.9\\
		$ \varepsilon $ in $ \varepsilon $-greedy algorithm & 0.1 to 0.005\\
		Learning rate used in RMSProp & 0.01\\
		Target network update frequency $ F $ & 200\\
		Experience-replay minibatch size $ N_E $ & 32\\
		Experience-replay memory capacity & 500\\
		\bottomrule
	\end{tabular}
\end{table}
\begin{figure*}[htbp]
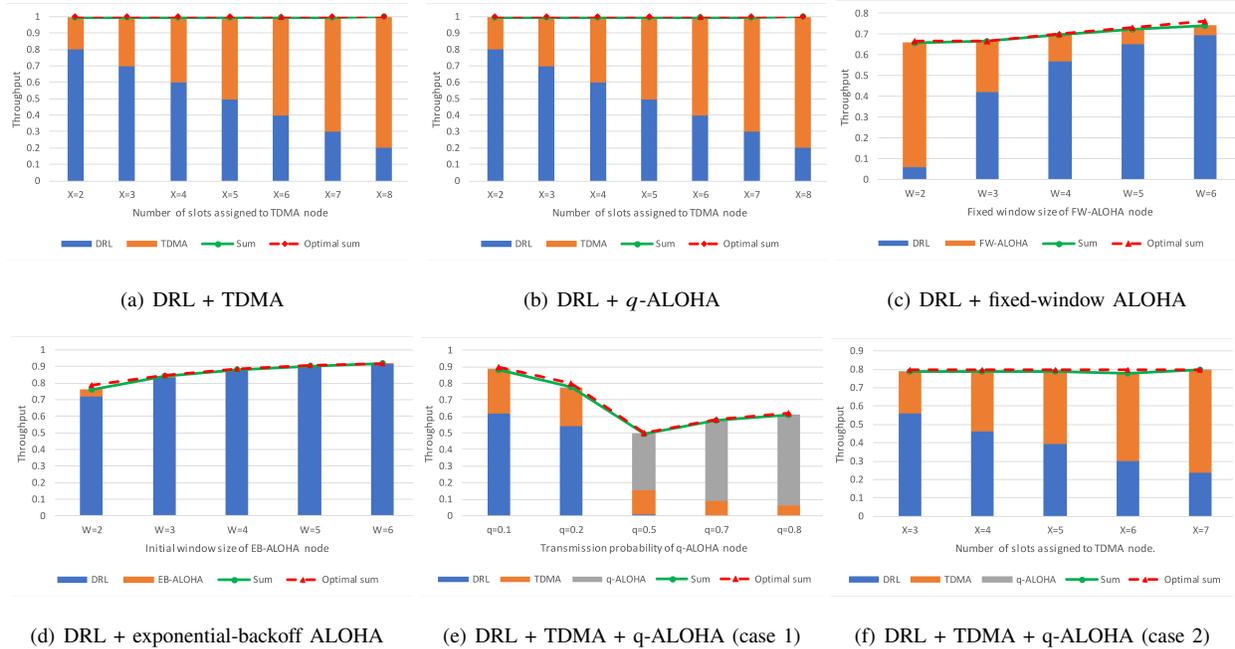

	\centering
	\begin{minipage}[t]{0.98\linewidth}
		\centering
		\subfigure[DRL + TDMA]{
			\begin{minipage}[t]{0.3\linewidth}
				\centering
				\includegraphics[width=1\linewidth]{agent+TDMA1.pdf}
				\label{fig:agent+TDMA1}
		\end{minipage}}
		\subfigure[DRL + $ q $-ALOHA]{
			\begin{minipage}[t]{0.3\linewidth}
				\centering
				\includegraphics[width=1\linewidth]{agent+TDMA1.pdf}
				\label{fig:agent+qALOHA1}
		\end{minipage}}
		\subfigure[DRL + fixed-window ALOHA]
		{\begin{minipage}[t]{0.3\textwidth}
				\centering
				\includegraphics[width=1\linewidth]{agent+FW-ALOHA1.pdf}
				\label{fig:agent+FW-ALOHA1}
		\end{minipage}}
		\subfigure[DRL + exponential-backoff ALOHA]{
			\begin{minipage}[t]{0.3\textwidth}
				\centering
				\includegraphics[width=1\linewidth]{agent+EB-ALOHA1.pdf}
				\label{fig:agent+EB-ALOHA1}
		\end{minipage}}
		\subfigure[DRL + TDMA + q-ALOHA (case 1)]
		{\begin{minipage}[t]{0.3\textwidth}
				\centering
				\includegraphics[width=1\linewidth]{agent+TDMA+qALOHA1_1.pdf}
				\label{fig:agent+TDMA+qALOHA1_1}
		\end{minipage}}
		\subfigure[DRL + TDMA + q-ALOHA (case 2)]
		{\begin{minipage}[t]{0.3\textwidth}
				\centering
				\includegraphics[width=1\linewidth]{agent+TDMA+qALOHA2_1.pdf}
				\label{fig:agent+TDMA+qALOHA2_1}
		\end{minipage}}
		\caption{Sum and individual throughputs for different cases with the objective of maximizing sum throughput.}
		\label{fig:six1}
		\vspace{-1em}
	\end{minipage}
\end{figure*}

A salient feature of our DLMA framework is that it is model-free (it does not need to know the protocols adopted by other coexisting nodes). For benchmarking, we consider model-aware nodes. Specifically, a model-aware node knows the MAC mechanisms of coexisting nodes, and it executes an optimal MAC protocol derived from this knowledge. We will show that our model-free DRL node can achieve near-optimal throughput with respect to the optimal throughput of the model-aware node. The derivations of the optimal throughputs for different cases below, which are interesting in their own right, are provided in \cite{benchmark}. We omit them here to save space. 
%\begin{figure*}[htbp]
%\centering
%\begin{minipage}{0.32\textwidth}
%\includegraphics[width=1\linewidth]{figs/agent+TDMA1.png}
%\caption{Throughputs under coexistence of one DRL node with one TDMA node.}
%\label{fig:agent+TDMA1}
%\end{minipage}
%\begin{minipage}{0.32\textwidth}
%\includegraphics[width=1\linewidth]{figs/agent+qALOHA1.png}
%\caption{Throughputs under coexistence of one DRL node with one $ q $-ALOHA node.}
%\label{fig:agent+qALOHA1}
%\end{minipage}
%\begin{minipage}{0.32\textwidth}
%\includegraphics[width=1\linewidth]{figs/agent+FW-ALOHA1.png}
%\caption{Throughputs under coexistence of one DRL node with one fixed-window ALOHA node.}
%\label{fig:agent+FW-ALOHA1}
%\end{minipage}
%\end{figure*}
%\begin{figure*}[htbp]
%\centering
%\begin{minipage}{0.32\textwidth}
%\includegraphics[width=1\linewidth]{figs/agent+EB-ALOHA1.png}
%\caption{Throughputs under coexistence of one DRL node with one exponential backoff ALOHA node.}
%\label{fig:agent+EB-ALOHA1}
%\end{minipage}
%\begin{minipage}{0.32\textwidth}
%\includegraphics[width=1\linewidth]{figs/agent+TDMA+qALOHA1_1.png}
%\caption{Throughputs under coexistence of one DRL node with one TDMA node and one $ q $-ALOHA node (the first case).}
%\label{fig:agent+TDMA+qALOHA1_1}
%\end{minipage}
%\begin{minipage}{0.32\textwidth}
%\includegraphics[width=1\linewidth]{figs/agent+TDMA+qALOHA2_1.png}
%\caption{Throughputs under coexistence of one DRL node with one TDMA node and one $ q $-ALOHA node (the second case).}
%\label{fig:agent+TDMA+qALOHA2_1}
%\end{minipage}
%%\vspace{-0.22in}
%\end{figure*}

\subsection{Coexistence with TDMA networks}
We first present the results of the coexistence of one DRL node with one TDMA node. The TDMA node transmits in $ X $  specific slots within each frame of $ Y $ slots in a repetitive manner from frame to frame. For benchmarking, we consider a TDMA-aware node which has full knowledge of the  $ X $ slots used by the TDMA node. To maximize the overall system throughput, the TDMA-aware node will transmit in all the $ Y-X $  slots not used by the TDMA node. The optimal sum throughput is one packet per time slot. The DRL agent, unlike the TDMA-aware node, does not know that the other node is a TDMA node (as a matter of fact, it does not even know how many other nodes there are) and just uses the DRL algorithm to learn the optimal strategy.

Fig. \ref{fig:agent+TDMA1} presents the throughput\footnote{Unless stated otherwise, ``throughput'' in this paper is the ``short-term throughput'', calculated as  $\sum\nolimits_{\tau  = t - N + 1}^t {{r_\tau }/N} $, where  $ N=1000 $. If one time step is 1 $ ms $ in duration, then this is the throughput over the past second. In the bar charts presented in this paper, ``throughput'' is the average reward over the last $ N $ steps in an experiment with a length of 50000 steps and we take the average of 10 experiments for each case to get the final value.} results when  $ Y=10 $ and $ X $ varies from 2 to 8. The green line is the sum throughput of the DRL node and the TDMA node. We see that it is very close to 1. This demonstrates that the DRL node can capture all the unused slots without knowing the TDMA protocol adopted by the other node. 
\subsection{Coexistence with ALOHA networks}
We next present the results of the coexistence of one DRL node with one  $ q $-ALOHA, one FW-ALOHA and one EB-ALOHA, respectively. We emphasize that the exact same DLMA algorithm as in Part A is used here even though the other protocols are not TDMA anymore.  For benchmarking, we consider model-aware nodes that operate with optimal MACs tailored to the operating mechanisms of the three ALOHA variants \cite{benchmark}.

Fig. \ref{fig:agent+qALOHA1} presents the experimental results for the coexistence of one DRL node and one  $ q $-ALOHA node. The results show that the DRL node can learn the strategy to achieve the optimal throughputs despite the fact that it is not aware that the other node is  $ q $-ALOHA node and what the transmission probability $ q $ is. Fig. \ref{fig:agent+FW-ALOHA1} presents the results for the coexistence of one DRL node and one FW-ALOHA node  with different fixed-window sizes. Fig. \ref{fig:agent+EB-ALOHA1} presents the results  for the coexistence of one DRL node and one EB-ALOHA node with different initial window sizes and maximum backoff stage $ m=2 $. As shown, DRL node can again achieve near-optimal throughputs for these two cases. 
\subsection{Coexistence with a mix of TDMA and ALOHA networks}\label{mix1}
We now present the results of a set-up in which one DRL agent coexists with one TDMA node and one  $ q $-ALOHA node simultaneously. Again, the same DLMA algorithm is used. We consider two cases. In the first case, the TDMA node transmits in 3 slots out of 10 slots in a frame; the transmission probability  $ q $ of the  $ q $-ALOHA node varies. In the second case,  $ q $ of the  $ q $-ALOHA node is fixed to 0.2;  $ X $, the number of slots used by the TDMA nodes in a frame, varies. Fig. \ref{fig:agent+TDMA+qALOHA1_1} and Fig. \ref{fig:agent+TDMA+qALOHA2_1} present the results of the first and second cases respectively. For both cases, we see that our DRL node can approximate the optimal results without knowing the transmission schemes of the TDMA and  $ q $-ALOHA nodes. 
\begin{figure}[!t]
	\centering
	\includegraphics[scale=0.48]{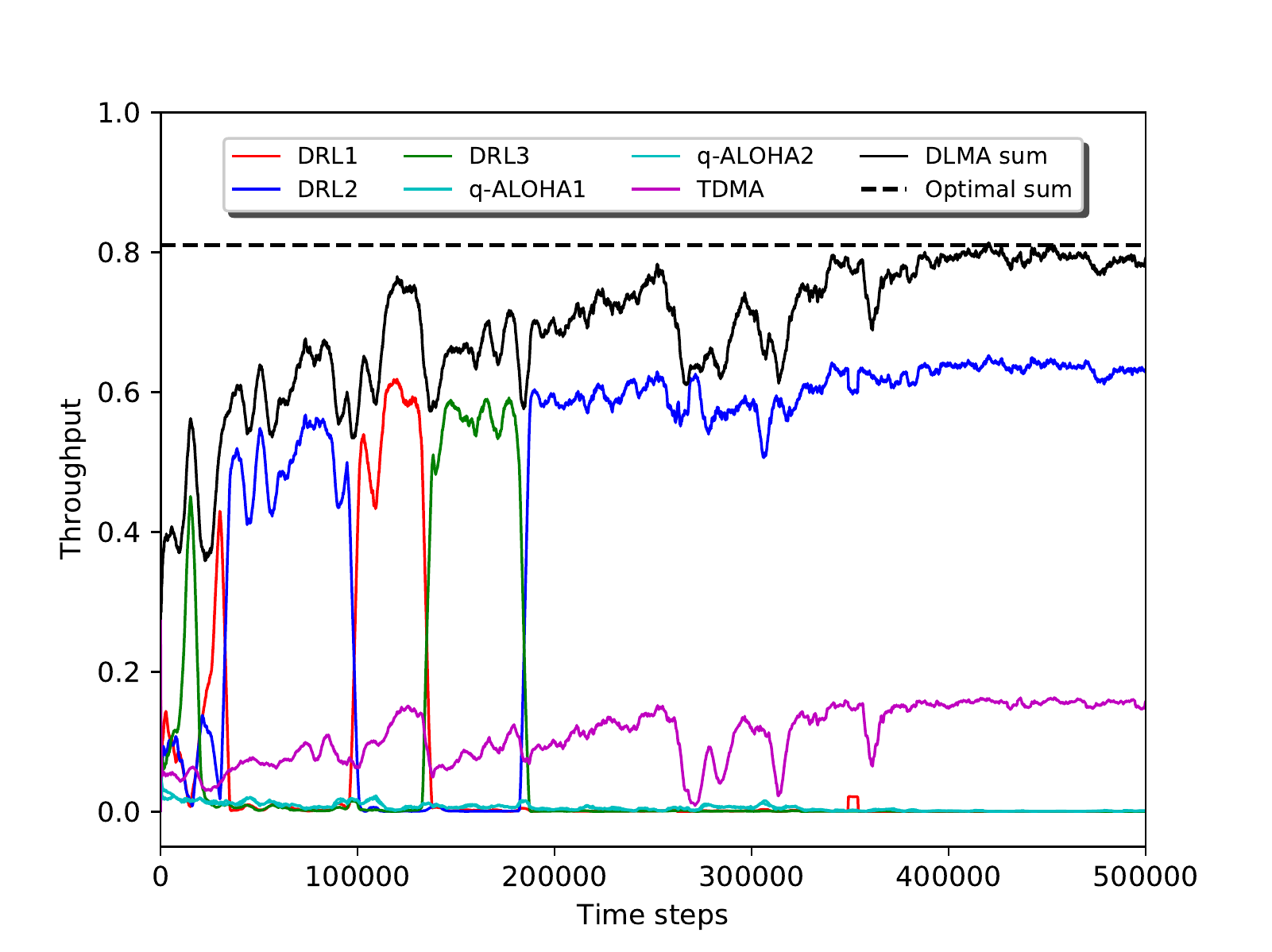}
	\caption{Sum throughput and individual throughputs under the coexistence of three DRL nodes with one TDMA node and two $  q $-ALOHA nodes. The throughputs at time $ t $ are computed by  $\sum\nolimits_{\tau  = t - N + 1}^t {{r_\tau }/N} $, where   $ N=5000 $.}
	\label{fig:sum}
\end{figure}

We next consider a setup in which multiple DRL nodes coexist with a mix of TDMA and  $ q $-ALOHA nodes. Specifically, the setup consists of three DRL nodes, one TDMA node that transmits in 2 slots out of 10 slots in a frame, and two  $ q $-ALOHA nodes with transmission probability  $ q=0.1 $. In Fig. \ref{fig:sum}, we can see that DLMA can also achieve near-optimal sum throughput in this more complex setup. However, when we focus on the individual throughputs of each node, we find that since there is no coordination among the three DRL nodes, one DRL node may preempt all the slots not occupied by the TDMA node, causing the other two DRL nodes and two q-ALOHA nodes get zero throughputs. This observation motivates us to consider fairness among different nodes in Section \ref{general_objective}.  

%\begin{figure*}[htbp]
%	\centering
%	\begin{minipage}{0.32\textwidth}
%		\includegraphics[width=1\linewidth]{figs/RLvsDRL1.png}
%		\caption{Convergence speeds of RL and DRL nodes. The sum throughput at time $ t $  is computed by   $\sum\nolimits_{\tau  = 1}^t {{r_\tau }/t} $.}
%		\label{fig:RLvsDRL1}
%	\end{minipage}
%	\begin{minipage}{0.32\textwidth}
%		\includegraphics[width=1\linewidth]{figs/RLvsDRL2.png}
%		\caption{Throughput evolution and cumulative number of visits to state  $ s_t $  prior to time   $ t $,  $ f\left( s_t, t\right)  $  , where  $ s_t $  is the particular state being visited at time  , for RL+TDMA and for DRL+TDMA. The sum throughput at time $ t $  is computed by  $\sum\nolimits_{\tau  = t - N + 1}^t {{r_\tau }/N} $, where   $ N=1000 $.}
%		\label{fig:RLvsDRL2}
%	\end{minipage}
%	\begin{minipage}{0.32\textwidth}
%		\includegraphics[width=1\linewidth]{figs/RLvsDRL3.png}
%		\caption{Total number of states visited by RL/DRL agent.}
%		\label{fig:RLvsDRL3}
%	\end{minipage}
%\end{figure*}

\subsection{RL versus DRL}\label{RLvsDRL}
\begin{figure}[!t]
	\centering
	\includegraphics[scale=0.38]{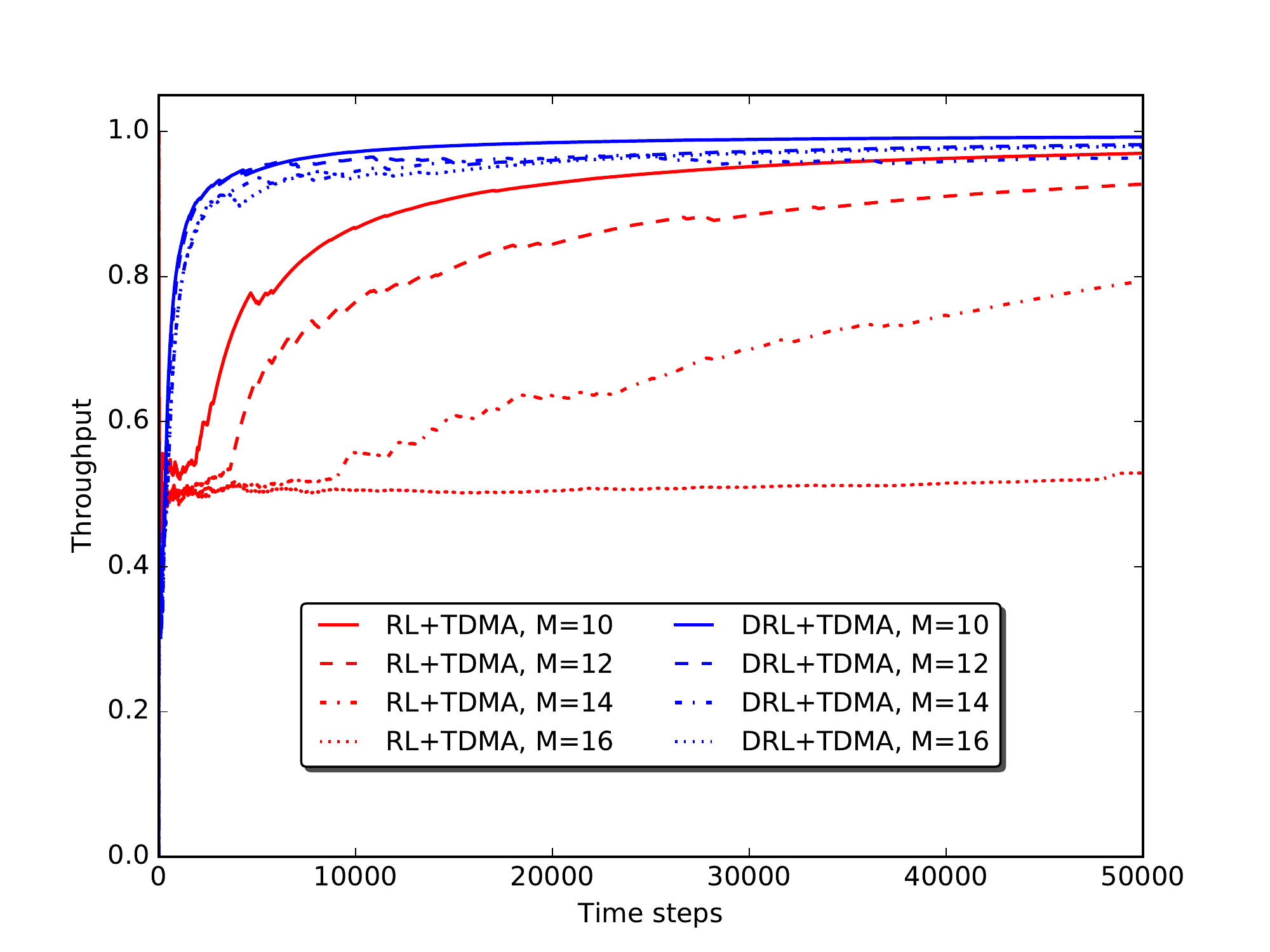}
	\caption{Convergence speeds of RL and DRL nodes. The sum throughput at time $ t $  is computed by   $\sum\nolimits_{\tau  = 1}^t {{r_\tau }/t} $.}
	\label{fig:RLvsDRL1}
\end{figure}

We now present results demonstrating the advantages of the ``deep'' approach using the scenario where one DRL/RL agent coexists with one TDMA node. Fig. \ref{fig:RLvsDRL1} compares the convergence time of the Q-learning based RL approach and the QNN-based DRL approach. The sum throughput in the figure is the ``cumulative sum throughput'' starting from the beginning:  $ \sum\nolimits_{\tau  = 1}^t {{r_\tau }} /t $. It can be seen that DRL converges to the optimal throughput of 1 at a much faster rate than RL does. For example, DRL requires only less than 5000 steps (5 $ s $ if each step corresponds to a packet transmission time of 1 $ ms $) to approach within 80$\%$ of the optimal throughput. Note that when state history length  increases from 10 to 16, RL learns progressively slower and slower, but the convergence time of DRL varies only slightly as $ M $ increases. In general, for a model-free MAC protocol, we do not know what other MAC protocols there are besides our MAC protocol. Therefore, we will not optimize on $M$ and will likely use a large $M$  to cater for a large range of other possible MAC protocols. The robustness, in terms of insensitivity of convergence time to $M$, is a significant practical advantage of DRL. 
\begin{figure}[!t]
	\centering
	\includegraphics[scale=0.38]{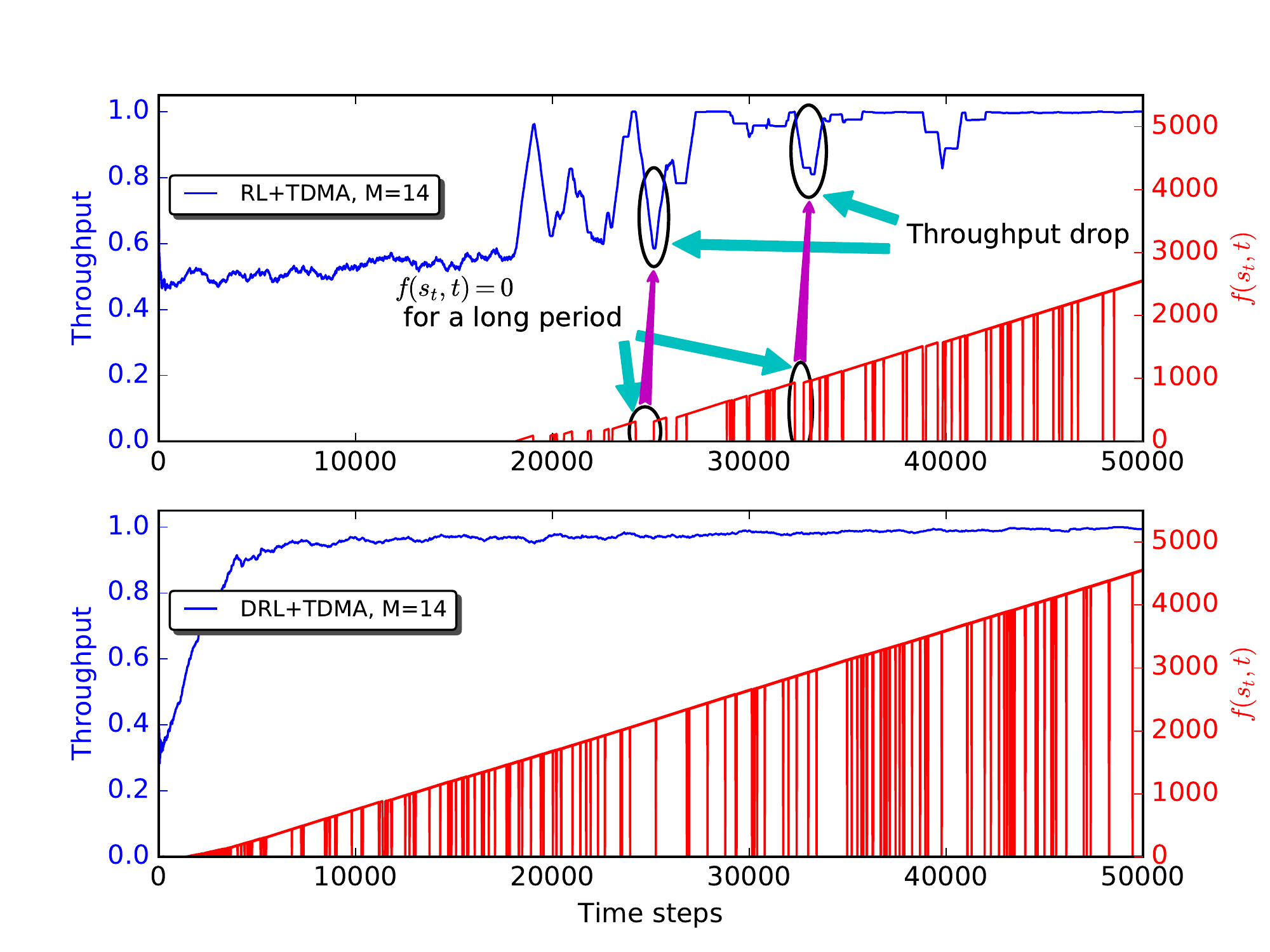}
	\caption{Throughput evolution and cumulative number of visits to state  $ s_t $  prior to time $ t $,  $ f\left( s_t, t\right)  $, where  $ s_t $  is the particular state being visited at time  , for RL+TDMA and for DRL+TDMA. The sum throughput at time $ t $  is computed by  $\sum\nolimits_{\tau  = t - N + 1}^t {{r_\tau }/N} $, where   $ N=1000 $. }
	\label{fig:RLvsDRL2}
	%\vspace{0.05in}
\end{figure}
\begin{figure}[!t]
	\centering
	\includegraphics[scale=0.48]{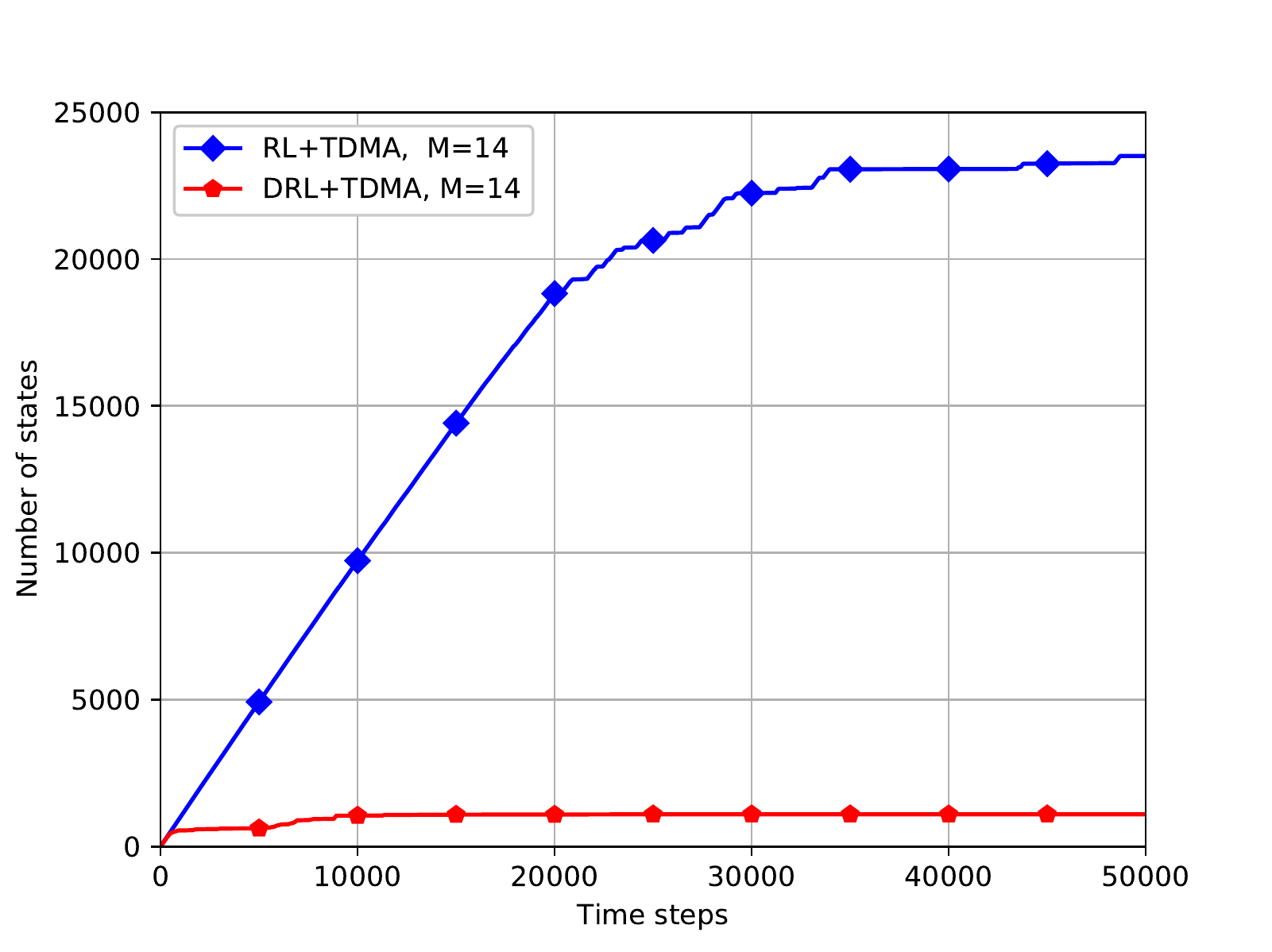}
	\caption{The number of distinct states visited by RL/DRL agent.}
	\label{fig:RLvsDRL3}
	%\vspace{0.01in}
\end{figure}
\begin{figure*}[!t]
	\subfigure[DRL + TDMA]{
		\label{fig:ResNet1}
		\begin{minipage}[t]{0.35\textwidth}
			\centering
			\includegraphics[scale=0.4]{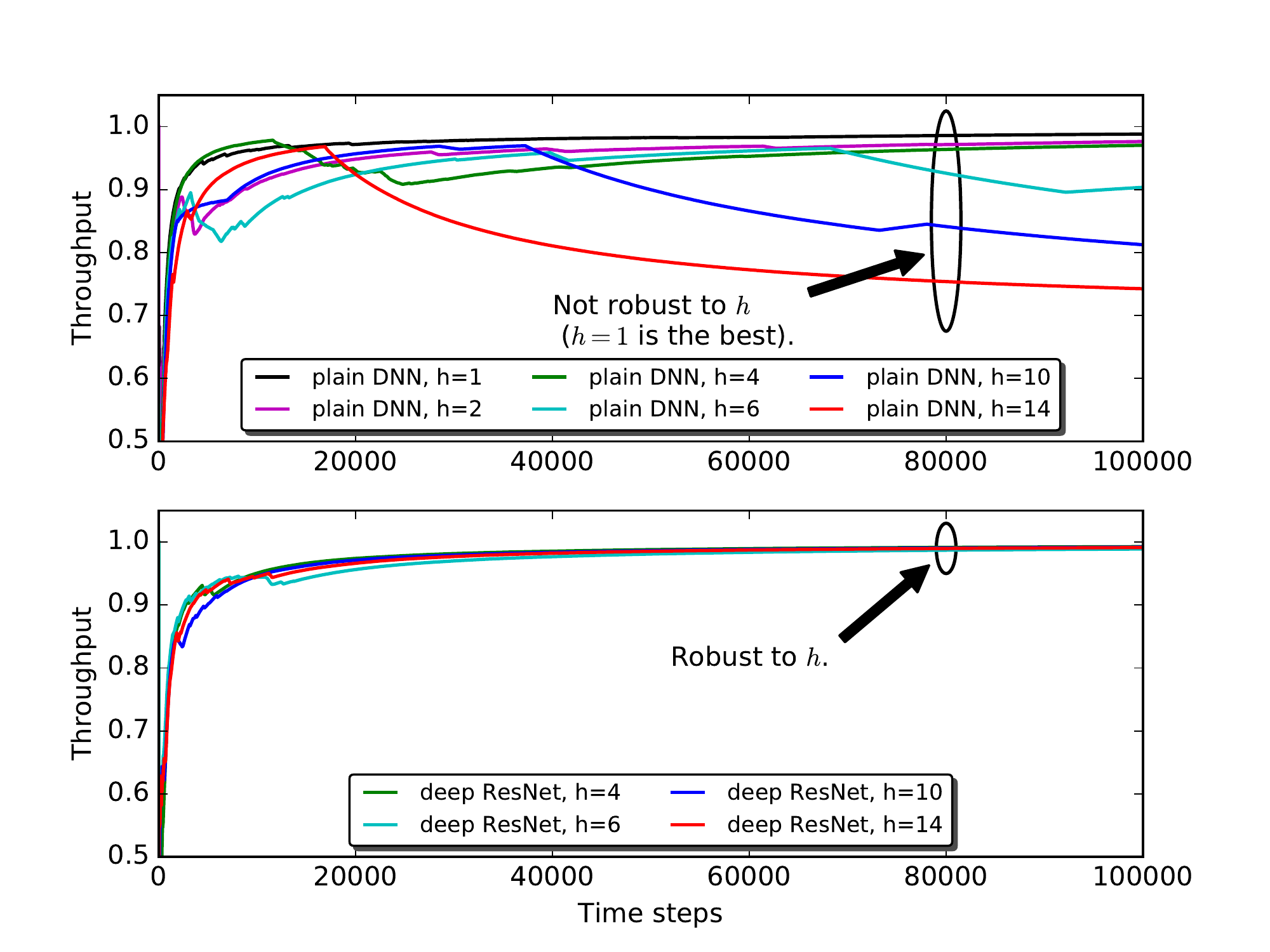}
	\end{minipage}}
	\hspace{2.5cm}
	\subfigure[DRL + TDMA + q-ALOHA]{
		\label{fig:ResNet2}
		\begin{minipage}[t]{0.35\textwidth}
			\centering
			\includegraphics[scale=0.4]{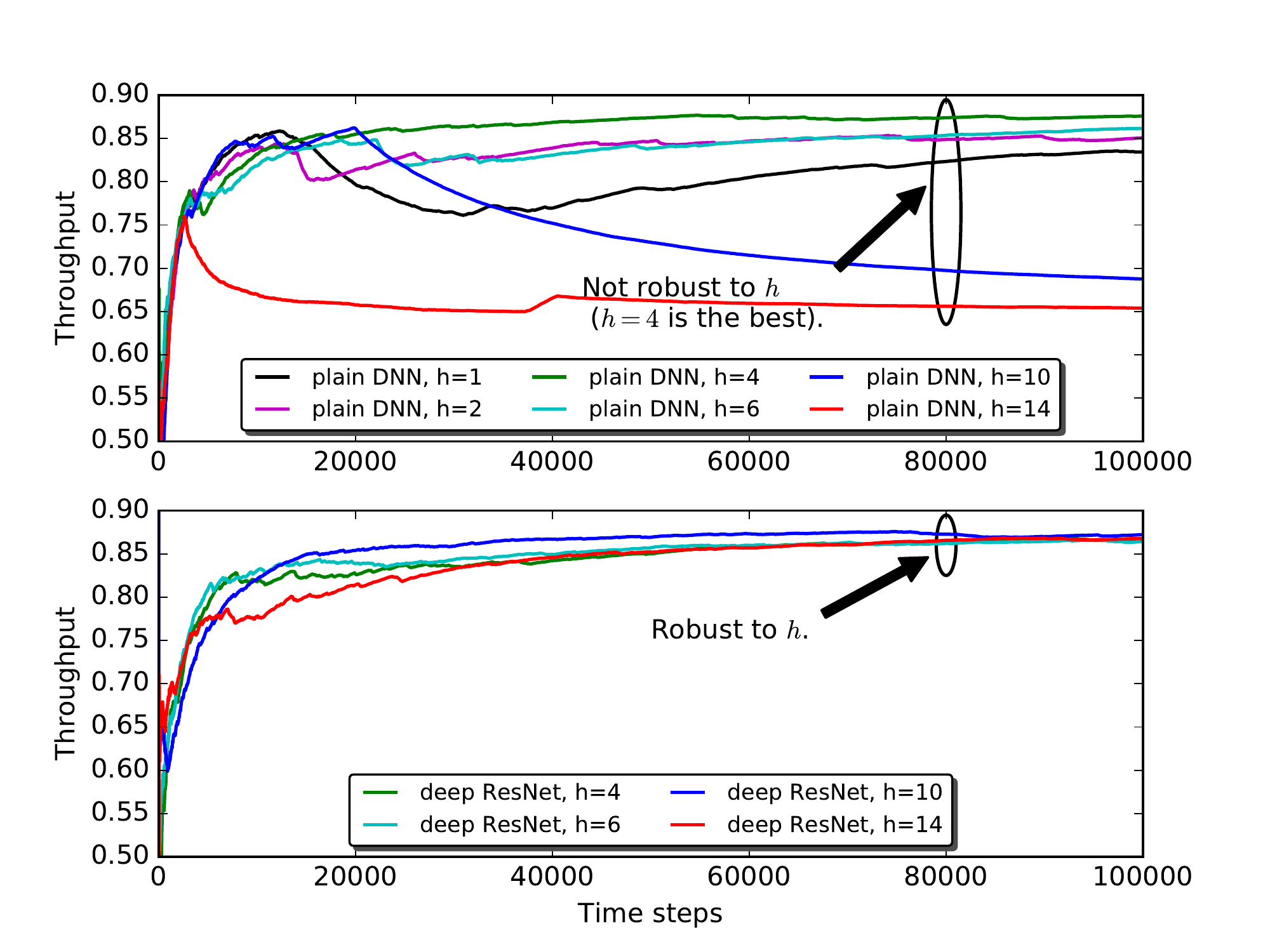}
	\end{minipage}}
	\caption{Performance comparison between plain DNN and deep ResNet based approaches. The cumulative sum throughput at time $ t $ is computed by  $\sum\nolimits_{\tau  = 1}^t {{r_\tau }/t} $.}
	\label{fig:ResNet}
\end{figure*}

Fig. \ref{fig:RLvsDRL2} presents the throughput evolutions of TDMA+RL and TDMA+DRL versus time. Unlike in Fig. \ref{fig:RLvsDRL1}, in Fig. \ref{fig:RLvsDRL2}, the sum throughput is the ``short-term sum throughput'' rather than the ``cumulative sum throughput'' starting from the beginning. Specifically, the sum throughput in Fig. \ref{fig:RLvsDRL2} is  $ \sum\nolimits_{\tau  = t - N + 1}^t {{r_\tau }} /N $, where $ N=1000 $. If one time step is 1 $ ms $ in duration, then this is the throughput over the past second. As can be seen, although both RL and DRL can converge to the optimal throughput in the end, DRL takes a much shorter time to do so. Furthermore, the fluctuations in throughput experienced by RL along the way are much larger. To dig deeper into this phenomenon, we examine $ f\left( s, t\right)  $, defined to be the number of previous visits to state $ s $ prior to time step  $ t $. Fig. \ref{fig:RLvsDRL2} also plots  $ f\left( s_t, t\right)  $: i.e., we look at the number of previous visits to state $ s_t $  before visiting $ s_t $, the particular state being visited at time step $ t $. As can be seen, for RL, each drop in the throughput coincides with a visit to a state  $ s_t $  with $ f\left( s_t, t\right)=0  $. In other words, the RL algorithm has not learned the optimal action for this state yet because of the lack of prior visits. From Fig. \ref{fig:RLvsDRL2}, we also see that it takes a while before RL extricates itself from persistent and consecutive visits to a number of states with $ f\left( \cdot \right)=0  $. This persistency results in large throughput drops until RL extricates itself from the situation.  By contrast, although DRL also occasionally visits a state $ s_t $  with $ f\left( s_t, t\right)=0  $, it is able to take an appropriate action at the unfamiliar territory (due to the ``extrapolation'' ability of the neural network to infer a good action to take at $ s_t $  based on prior visits to states other than  $ s_t $: recall that each update of $\bm{\theta}$ changes the values of  $q({\kern 1pt} s,a,\bm{\theta} {\kern 1pt} )$ for all  $\left( s, a\right)  $, not just that of a particular $\left( s, a\right)  $). DRL manages to extricate itself from unfamiliar territories quickly and evolve back to optimal territories where it only transmits in time slots not used by TDMA. 

Fig. \ref{fig:RLvsDRL3} presents the evolutions of the number of distinct states visited by RL and DRL agents in the same experiment as in Fig. \ref{fig:RLvsDRL2}. In this case, both RL and DRL find an optimal strategy in the end, but RL requires more time to do so. Once the optimal strategies are found, RL and DRL agents seldom explore new states, except the ``exploration'' step in $ \varepsilon $-greedy algorithm. As indicated in Fig. \ref{fig:RLvsDRL3}, the number of distinct states visited by RL on its journal to the optimal strategy is much larger than that of DRL. From Fig. \ref{fig:RLvsDRL3}, we  see that RL spends 35000 time steps in finding the optimal strategy, having visited 23000 distinct states before doing so. By contrast, it takes only 10000 time steps for DRL to find the optimal strategy and the number of distinct states visited is only around 1000. In other words, DRL can better narrow down its choice of states to visit in order to find the optimal strategy, hence the faster convergence speed. 

\subsection{Plain DNN versus deep ResNet}\label{sec:ResNet}
We now demonstrate the advantages of deep ResNet over plain DNN using two cases: 1) one DRL node coexisting with one TDMA node, wherein the TDMA node occupies 2 slots out of 10 slots in a frame; 2) one DRL node coexisting with one TDMA node and one q-ALOHA node, wherein the TDMA is the same as in 1) and  $ q=0.1 $ for the q-ALOHA node. The optimal sum throughputs for a model-aware protocol  for 1) and 2) can be established analytically to be  1 and 0.9, respectively (see \cite{benchmark} for the derivation). For each case, we compare the cumulative sum throughputs of plain DNN based approach and deep ResNet based approach with different numbers of hidden layers $ h $. 

As can be seen from the upper parts of Fig. \ref{fig:ResNet1} and Fig. \ref{fig:ResNet2}, the plain DNN is not robust against variation of   $ h $, i.e., the performance varies with   $ h $. Furthermore, $ h=1 $ and  $ h=4 $  achieve the best performance for case 1) and 2), respectively. This implies that it is difficult to use a common plain DNN architecture for different wireless setups.  In other words, the optimal $ h $ may be different under different scenarios. If the environment changes dynamically, there is no one single $ h $ that is optimal for all scenarios. In contrast to plain DNN's non-robustness to $ h $, deep ResNet can always achieve near-optimal performance for different $ h $  for both cases, as illustrated in the lower parts of Fig. \ref{fig:ResNet1} and Fig. \ref{fig:ResNet2}. 

For wireless networking, the environment may change quickly when new nodes arrive, and existing nodes move or depart. It is desirable to adopt a one-size-fits-all neural network architecture in DRL. Our results show that deep ResNet is more desirable than plain DNN in this regard.
  
%\begin{figure}[!t]
%\centering
%\includegraphics[scale=0.4]{figs/ResNetvsFNN1.png}
%\caption{Performance comparison between DNN and ResNet based approaches. The sum throughput at time $ t $ is computed by  $\sum\nolimits_{\tau  = 1}^t {{r_\tau }/t} $.}
%\label{fig:ResNetvsFNN}
%\vspace{-0.22in}
%\end{figure}

\section{General Objective DLMA protocol}\label{general_objective}
This section first introduces the well-known  $ \alpha $-fairness utility function \cite{mo2000fair}. Then, a multi-dimensional Q-learning algorithm is proposed to incorporate the  $ \alpha $-fairness utility function in a general reformulation of DLMA. 
\subsection{$ \alpha $-fairness objective}
Instead of sum throughput, we now adopt the  $ \alpha $-fairness index as the metric of the overall system performance. The parameter $ \alpha  \in [0,\infty ) $  is used to specify a range of the fairness criteria, e.g., when $ \alpha=0 $ , maximizing the $ \alpha $-fairness objective corresponds to maximizing the sum throughput (the corresponding results were presented in Section \ref{exp_sum}); when $ \alpha=1 $, maximizing the $ \alpha $-fairness objective corresponds to achieving proportional fairness; when  $ \alpha\rightarrow \infty $, the minimum throughput among nodes is being maximized. Specifically, we consider a system with $ N $  nodes and for a particular node  $ i $, its throughput is denoted by  $ x^{\left( i\right) } $; its  $ \alpha $-fairness local utility function is given by
\begin{equation}
f_\alpha ^{\left( i \right)}\left( {{x^{\left( i \right)}}} \right) = \left\{ {\begin{array}{*{20}{c}}
{\log \left( {{x^{\left( i \right)}}} \right)},\quad & \text{if} \quad \alpha = 1,\\
{{{\left( {1 - \alpha } \right)}^{ - 1}}{{\left( {{x^{\left( i \right)}}} \right)}^{1 - \alpha }}},\quad & \text{if} \quad \alpha  \ne 1.
\end{array}} \right.
\end{equation}
The objective of the overall system is to maximize the sum of all the local utility functions:
\begin{align}\label{fair-objective}
\text{maximize} \quad &F\left( {{x^{\left( 1 \right)}},{x^{\left( 2 \right)}}, \ldots ,{x^{\left( N \right)}}} \right) = \sum\limits_{i = 1}^N {f_\alpha ^{\left( i \right)}\left( {{x^{\left( i \right)}}} \right)} \nonumber \\
\text{subject to} \quad 
&\sum\nolimits_{i = 1}^N {{x^{\left( i \right)}}}  \le 1, \\
&{x^{\left( i \right)}} \ge 0,\quad \forall i. \nonumber
\end{align}
\subsection{DLMA reformulation}
We now reformulate our system model as a semi-distributed system that consists of several wireless networks with different MAC protocols.  Nodes in different networks cannot communicate with each other. For nodes within the DLMA network, there is a DLMA central gateway that coordinates the transmissions of the nodes. Similarly, for nodes within the TDMA network, there is implicitly a TDMA central gateway to decide the time slots in which TDMA nodes transmit. 

Among the  $ N $  nodes in the wireless networks, let $ K $  be the number of DRL nodes in the DLMA network and  $ L=N-K $ be the number of non-DRL nodes. In the DLMA protocol as described in Section \ref{DLMA_formulation},  all DRL nodes individually adopt the single-agent DRL algorithm, and independently perform training and execution of the DRL algorithm.  Unlike the DLMA protocol in Section \ref{DLMA_formulation}, we now consider an DRL algorithm with ``centralized training at the gateway node and independent execution at DRL nodes''. The gateway in the DLMA network associates with all other DRL nodes in the DLMA network and coordinates the coexistence of the DLMA network with other networks (e.g., the TDMA and ALOHA networks).  In each time slot, the gateway decides whether a node in the DLMA network should transmit or not. If YES,  the gateway selects one of the DRL nodes in a round-robin manner to transmit. After transmitting, the selected DRL node receives a feedback from the system and communicates with the gateway with this information. If NO, all DRL nodes keep silent. In this manner, the  gateway can be regarded as a virtual big agent that is a combination of  the $ K $  DRL nodes. The coordination information from the gateway  to other DRL nodes can be sent through a control channel. For example, the control channel can be implemented as a short time slot after each time slot of information transmission. Other implementations are also possible, but we will omit the discussion here since the focus of this paper is not implementation details. The above reformulates the system to contain  $ L+1 $ nodes: one DRL big agent node (we index it by  $ i=L+1 $)  and other $ L $  legacy nodes (we index them by  $ i=1,2,\cdots, L $). 

We now modify the original Q-learning algorithm. The original Q-learning algorithm is designed for the single-agent case under the objective of maximizing the accumulated reward of the agent. It cannot be directly applied to the multi-node/multi-agent  case to meet arbitrary fairness objectives. We therefore put forth a multi-dimensional Q-learning algorithm to cater for the  $ \alpha $-fairness objective.  

In the original Q-learning algorithm, each agent receives a scalar reward from the environment. The scalar reward, representing the overall system transmission result (success, idleness or collision), is regarded as the overall reward to the system in the original Q-learning algorithm. Each agent uses the overall reward to compute the sum throughput objective. By contrast, in our new multi-dimensional Q-learning algorithm, the big agent receives an   $ L+1 $ dimension vector of rewards from the environment. Each element of the vector represents the transmission result of one particular node. The reward vector is used to compute the  $ \alpha $-fairness objective. Specifically, let $ {r^{\left( i \right)}} $  be the reward of node $ i $  and thus the received reward vector is given by   $  \left[ {{r^{\left( i \right)}}} \right]_{i = 1}^{L + 1}$. For a state-action pair  $ (s,a) $, instead of maintaining an action-value scalar  $Q\left( {s,a} \right)$, the big agent maintains an action-value vector  $\left[ {{Q^{\left( i \right)}}\left( {s,a} \right)} \right]_{i = 1}^{L + 1}$, where  the element ${Q^{\left( i \right)}}\left( {s,a} \right)$  is the expected accumulated discounted reward of node  $ i $. 

Let  $ {q^{\left( i \right)}}\left( {s,a} \right) $ be the estimate of the elementary action-value function ${Q^{\left( i \right)}}\left( {s,a} \right)$  in the action-value vector. Suppose at time $ t $, the state is $ s_t $. For decision making, we still adopt the  $ \varepsilon $-greedy algorithm. When selecting the greedy action, the objective in (\ref{fair-objective}) can be applied to meet arbitrary fairness objective, i.e., 
\begin{align}\label{greedy_action}
{a_t} = \arg {\max _a} &\left\{ \sum\limits_{i = 1}^L {f_\alpha ^{\left( i \right)}\left( {{q^{\left( i \right)}}\left( {{s_t},a} \right)} \right)} + \right.\nonumber  \\
& \left. K \cdot f_\alpha ^{\left( {L + 1} \right)}\left( {\frac{{{q^{\left( {L + 1} \right)}}\left( {{s_t},a} \right)}}{K}} \right) \right\}.
\end{align}
After taking action $ a_t $, the big agent employs the multi-dimensional Q-learning algorithm to parallelly update the  $ L+1 $ elementary action-value estimates   ${q^{\left( i \right)}}({s_t},{a_t})$,  $i = 1,2, \ldots L + 1$ as
\begin{align}\label{Qupdate2}
{q^{\left( i \right)}}\left( {{s_t},{a_t}} \right) \leftarrow &{q^{\left( i \right)}}\left( {{s_t},{a_t}} \right) +  \nonumber \\
& \beta \left[ {r_{t + 1}^{\left( i \right)} + \gamma {q^{\left( i \right)}}\left( {{s_{t + 1}},{a_{t + 1}}} \right) - {q^{\left( i \right)}}\left( {{s_t},{a_t}} \right)} \right],
\end{align}
where
\begin{align}\label{greedy_action2}
{a_{t + 1}} = \arg {\max _{a'}} & \left\{ \sum\limits_{i = 1}^L {f_\alpha ^{\left( i \right)}\left( {{q^{\left( i \right)}}\left( {{s_{t + 1}},a'} \right)} \right)}  + 
\right.\nonumber  \\
& \left. K \cdot f_\alpha ^{\left( {L + 1} \right)}\left( {\frac{{{q^{\left( {L + 1} \right)}}\left( {{s_{t + 1}},a'} \right)}}{K}} \right) \right\}.
\end{align}

Here, it is important to point out the subtleties in (\ref{greedy_action})-(\ref{greedy_action2}) and how they differ from the conventional Q-learning update equation in (\ref{Qupdate}). In conventional Q-learning, an action that optimizes the Q function is chosen (as explained in Section \ref{RLoverview}). In other words, the Q function is the objective function to be optimized. However, the Q function as embodied in (\ref{Qupdate}) and (\ref{Qupdate2}) is a projected (estimated) weighted sum of the current rewards and future rewards. To be more specific, take a look at the term $ \left[ {{r_{t + 1}} + \gamma \mathop {\max }\limits_{a'} q\left( {{s_{t + 1}},a'} \right)} \right] $  in (3). It can be taken to be an estimation of   $[{r_{t + 1}} + \gamma E[{r_{t + 2}}] + {\gamma ^2}E[{r_{t + 3}}] + ...]$, which is a weighted sum of the current reward and future rewards with discount factor  $ \gamma $. We can view  $[{r_{t + 1}} + \gamma E[{r_{t + 2}}] + {\gamma ^2}E[{r_{t + 3}}] + ...]$ as a newly estimated  $ q\left( {{s_t},{a_t}} \right) $. In (\ref{Qupdate}), for the purpose of estimation smoothing, we apply a weight of  $ \beta $ to this new estimate and a weight of $(1 - \beta )$  to the previous value of $ q\left( {{s_t},{a_t}} \right) $ to come up with a new value for  $ q\left( {{s_t},{a_t}} \right) $. Nevertheless, $ q\left( {{s_t},{a_t}} \right) $ still embodies a weighted sum of the current and future rewards. Since in conventional Q-learning,  an action  that gives the maximum $ q\left( {{s_t},{a_t}} \right) $ is taken at each step  $ t $, the objective can be viewed as trying to maximize a weighted sum of rewards with discount factor  $ \gamma $. However, not all objectives can be conveniently expressed as a weighted sum of rewards. An example is the  $ \alpha $-fairness objective of focus here. A contribution of us here is the realization that, for generality, we need to separate the objective upon which the optimizing action is chosen and the Q function itself.
%\footnote{\textcolor{red}{This is true not just for Q-learning in the wireless network application here, but also true if we want to generalize and extend the application of Q-learning in other domains. As far as we know, we are the first to propose this generalization of Q learning. Furthermore, although we explain this generalization in terms of the multi-agent setting, it is also relevant to the single-agent setting whereby the single-agent may incur a number of different types of rewards, with the objective being a complex function of the different types of rewards. In such a case, each reward can be associated with a Q function, with the objective being a function of the underlying Q functions. }}  

Objectives can often be expressed as a function of several components,  wherein each component can be expressed as a Q function (e.g., (\ref{greedy_action})). In the more general setup, the update equation of Q function still has the same form (i.e., (\ref{Qupdate2}) has the same form as (\ref{Qupdate})). However, the action $ a_{t+1} $  chosen a time step later in (\ref{Qupdate2}) is not that gives the maximum  $ {q^{\left( i \right)}}\left( {{s_{t + 1}},{\kern 1pt} {\kern 1pt}  \cdot } \right) $, but which is based on (\ref{greedy_action2}). Thus, the Q function is still a projected weighted sum of rewards. But the policy that gives rise to the rewards is not based on maximizing the weighted sum of rewards, but based on maximizing a more general objective. 

Returning to our wireless setting, the first term in (\ref{greedy_action}) is the sum of local utility functions of all legacy nodes. Since  the big agent (indexed by   $ L+1 $) is actually a combination of the  $ K $ DRL nodes, and $ {q^{\left( {L + 1} \right)}}\left( \cdot \right)/K $  is the estimated accumulated reward of each DRL node, the second term in (\ref{greedy_action}) is the sum of local utility functions of all the DRL nodes. We have two remarks:  i) the $ {q^{\left( i \right)}}\left( {{s_t},{a_t}} \right) $  in (\ref{Qupdate2}) is an estimate of the expected accumulated discounted reward of node $ i $ (as expressed in (\ref{greedy_action}), rather than the exact throughput  $ {x^{\left( i \right)}} $  in (\ref{fair-objective}); ii) we use $ {q^{\left( i \right)}}\left( {{s_t},{a_t}} \right) $  to help the agent make decisions because the exact throughput $ {x^{\left( i \right)}} $  is not known. Our evaluation results in section \ref{exp_pro} show that this method can achieve the fairness objective. 
\begin{algorithm}[!t]
	\caption{DLMA with the $ \alpha $-fairness objective}\label{alg:DLMA2}
	\begin{algorithmic}
		\State Initialize $ s_0 $, $ \varepsilon $, $ \gamma $, $ \rho $,  $ N_E $, $ F $
		\State Initialize experience memory $ EM $
		\State Initialize the parameter of QNN as $  \bm{\theta } $ 
		\State Initialize the parameter of target QNN  $ \bm{\theta^- }=\bm{\theta } $ 
		\For{$ t=0,1,2,\cdots $ in DLMA}
		\State Input $ s_t $ to QNN and output  
		\begin{equation*}
		\mathbb{Q} = \left\{ {{q^{\left( i \right)}}\left( {s_t,a,{\bm{\theta }}} \right)|a \in {A_{s_t}}}, i=1,2,\cdots, L+1 \right\}
		\end{equation*}
		\State Generate action $ a_t $ from $ \mathbb{Q} $ using $ \varepsilon $-greedy algorithm
		\State Observe $ z_t $, $ r^{\left( 1 \right)}_{t+1} $ , $ r^{\left( 2 \right)}_{t+1} $, $ \cdots $, $ r^{\left( L \right)}_{t+1} $
		\State Compute $ s_{t+1} $ from $ s_t $, $ a_t $ and $ z_t $
		%\If{$ \vert EM \vert < \vert EM \vert_{max} $}
		%	\State Add $ e_t = \left( {s_t,a_t,{r^{\left( 1 \right)}_{t+1}},{r^{\left( 2 \right)}_{t+1}}, \ldots {r^{\left( {L + 1} \right)}_{t+1}},s_{t+1}} \right) $ to  $ EM $
		%\Else
		%	\State Add $ e_t $ to $ EM $ in an FIFO manner
		%\EndIf
		\State Store $ \left( {s_t,a_t,{r^{\left( 1 \right)}_{t+1}},{r^{\left( 2 \right)}_{t+1}}, \ldots {r^{\left( {L + 1} \right)}_{t+1}},s_{t+1}} \right) $ to  $ EM $
		\State \textbf{if} Remainder($ t/F==0 $) \textbf{then}  $ I=1 $ \textbf{else} $ I=0 $
		\State \Call{TrainQNN}{$ \gamma $, $ \rho $, $ N_E $, $ I $, $ EM $, $ \bm{\theta} $, $ \bm{\theta^-} $}
		\EndFor \\
		\Procedure{TrainQNN}{$ \gamma $, $ \rho $, $ N_E $, $ I $, $ EM $, $ \bm{\theta} $, $ \bm{\theta^-} $}
		\State Randomly sample $ N_E $ experience tuples from $ EM $ as $ E $
		\For{each sample $ e = \left( {s,a,r^{\left( 1 \right)}, r^{\left( 2 \right)}, \cdots, r^{\left( L+1 \right)},s'} \right) $ in $ E $}
		\State Calculate ${y^{\left( i \right)}}_{r,s'}^{QNN} = {r^{\left( i \right)}} + \gamma {q^{\left( i \right)}}\left( {s',a';{{\bm{\theta }}^ - }} \right)$, where $ a' $ is selected to according (\ref{greedy_action3})
		\EndFor 
		\State Perform Gradient Descent to update  $ \bm{\theta } $ in QNN:
		\begin{align*}
		\text{Iterate} \ {\bm{\theta }} \leftarrow &{\bm{\theta }} - \frac{\rho }{{{N_E}\left( {L + 1} \right)}} \cdot \\ &\sum\limits_{i = 1}^{L + 1} {\sum\limits_{e \in E} {\left[ {{y^{\left( i \right)}}_{r,s'}^{QNN} - {q^{\left( i \right)}}\left( {s,a;{\bm{\theta }}} \right)} \right]\nabla {q^{\left( i \right)}}\left( {s,a;{\bm{\theta }}} \right)} } 
		\end{align*}
		\If{$ I==1 $}
		\State Update $ \bm{\theta^- } $ in target QNN by setting $ \bm{\theta^-} = \bm{\theta } $
		\EndIf
		\EndProcedure
	\end{algorithmic}
\end{algorithm}
\begin{figure*}[htbp]
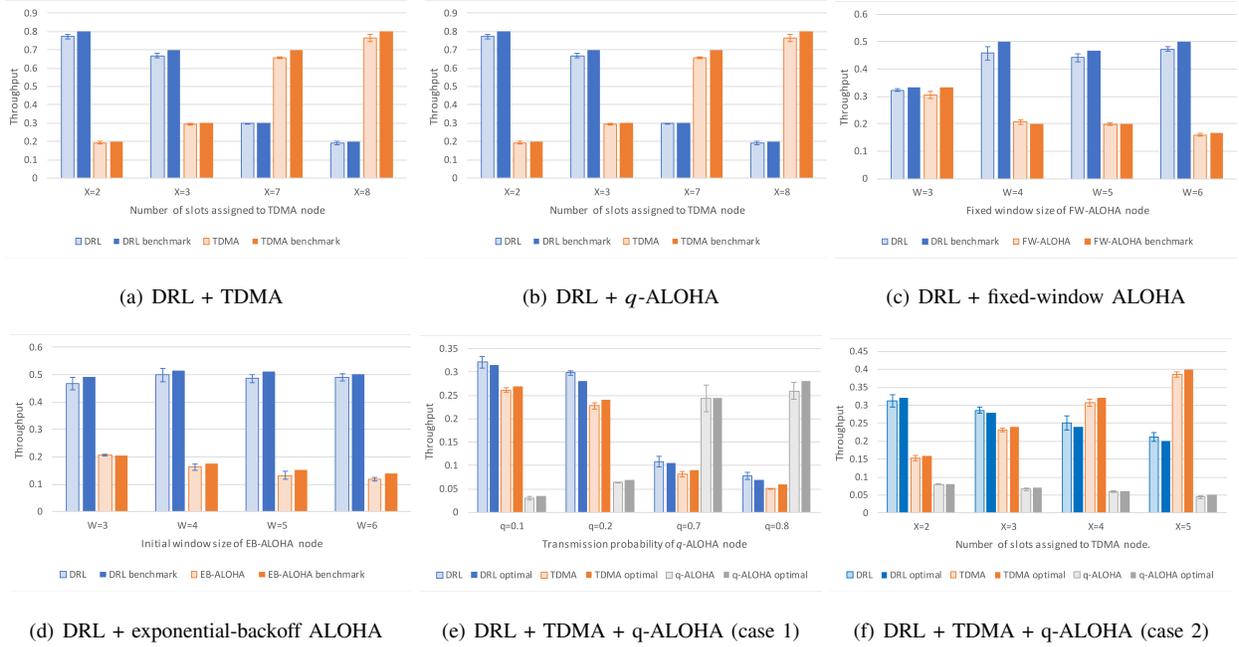

	\centering
	\begin{minipage}[t]{0.98\linewidth}
		\centering
		\subfigure[DRL + TDMA]{
			\begin{minipage}[t]{0.3\linewidth}
				\centering
				\includegraphics[width=1\linewidth]{agent+TDMA2.pdf}
				\label{fig:agent+TDMA2}
		\end{minipage}}
		\subfigure[DRL + $ q $-ALOHA]{
			\begin{minipage}[t]{0.3\linewidth}
				\centering
				\includegraphics[width=1\linewidth]{agent+TDMA2.pdf}
				\label{fig:agent+qALOHA2}
		\end{minipage}}
		\subfigure[DRL + fixed-window ALOHA]
		{\begin{minipage}[t]{0.3\textwidth}
				\centering
				\includegraphics[width=1\linewidth]{agent+FW-ALOHA2.pdf}
				\label{fig:agent+FW-ALOHA2}
		\end{minipage}}
		\subfigure[DRL + exponential-backoff ALOHA]{
			\begin{minipage}[t]{0.3\textwidth}
				\centering
				\includegraphics[width=1\linewidth]{agent+EB-ALOHA2.pdf}
				\label{fig:agent+EB-ALOHA2}
		\end{minipage}}
		\subfigure[DRL + TDMA + q-ALOHA (case 1)]
		{\begin{minipage}[t]{0.3\textwidth}
				\centering
				\includegraphics[width=1\linewidth]{agent+TDMA+qALOHA1_2.pdf}
				\label{fig:agent+TDMA+qALOHA1_2}
		\end{minipage}}
		\subfigure[DRL + TDMA + q-ALOHA (case 2)]
		{\begin{minipage}[t]{0.3\textwidth}
				\centering
				\includegraphics[width=1\linewidth]{agent+TDMA+qALOHA2_2.pdf}
				\label{fig:agent+TDMA+qALOHA2_2}
		\end{minipage}}
		\caption{Individual throughputs for different cases with the objective of achieving proportional fairness.}
		\label{fig:six2}
	\end{minipage}
\end{figure*}
We continue the reformulation of DLMA by incorporating deep neural networks. The incorporation of deep neural networks into the multi-dimensional Q-learning algorithm calls for two additional modifications. The first is to use a QNN to approximate the action-value vector $\left[ {{Q^{\left( i \right)}}\left( {s,a} \right)} \right]_{i = 1}^{L + 1}$  as  $ \left[ {{q^{\left( i \right)}}\left( {s,a;{\bm{\theta }}} \right)} \right]_{i = 1}^{L + 1} $, where $ \bm{\theta } $  is the weights of QNN. 
The second is to augment the experience tuple to  $ e = \left( {s,a,{r^{\left( 1 \right)}},{r^{\left( 2 \right)}}, \ldots {r^{\left( {L + 1} \right)}},s'} \right) $. With these two modifications, the loss function (\ref{loss2}), the target (\ref{yQNN2}) and the update of $ \bm{\theta} $ (\ref{theta_update2}) are now given by

	\begin{equation}\label{loss3}
	{L_E}\left( {{\bm{\theta }}} \right) = \frac{1}{{{N_E}\left( {L + 1} \right)}}\sum\limits_{i = 1}^{L + 1} {\sum\limits_{e \in {E_t}} {{{\left( {{y^{\left( i \right)}}_{r,s'}^{QNN} - {q^{\left( i \right)}}\left( {s,a;{\bm{\theta }}} \right)} \right)}^2}} },
	\end{equation}
	\begin{equation}\label{yQNN3}
	{y^{\left( i \right)}}_{r,s'}^{QNN} = {r^{\left( i \right)}} + \gamma {q^{\left( i \right)}}\left( {s',a';{\bm{\theta^- }}} \right),
	\end{equation}
	where
	\begin{align}\label{greedy_action3}
	a' = \arg {\max _{\tilde a'}} &\left\{ \sum \limits_{i = 1}^L {f_\alpha ^{\left( i \right)}\left( {{q^{\left( i \right)}}\left( {s',\tilde a';{{\bm{\theta^- }}}} \right)} \right)}  + \right. \nonumber \\
	& \left. K \cdot f_\alpha ^{\left( {L + 1} \right)}\left( {\frac{{{q^{\left( {L + 1} \right)}}\left( {s',\tilde a';{{\bm{\theta^- }}}} \right)}}{K}} \right) \right\},
	\end{align}
	\begin{align}\label{theta_update3}
	\text{Iterate} \ \bm{\theta} \leftarrow &{\bm{\theta}} - \frac{\rho}{N_E\left( L + 1\right)} \cdot \nonumber \\ &\sum\limits_{i = 1}^{L + 1} {\sum\limits_{e \in {E}} {\left[ {{y^{\left( i \right)}}_{r,s'}^{QNN} - {q^{\left( i \right)}}\left( {s,a;{\bm{\theta}}} \right)} \right]\nabla {q^{\left( i \right)}}\left( {s,a;{\bm{\theta}}} \right)} }.
	\end{align}
The pseudocode of the reformulated DLMA protocol is summarized in Algorithm \ref{alg:DLMA2}.

\section{Proportional Fairness Performance Evaluation}\label{exp_pro}
This section investigates the performance when DRL nodes aim to achieve proportional fairness among nodes, as a representative example of the general  $ \alpha $-fairness DLMA formulation. We investigate the interaction of DRL nodes with TDMA nodes, ALOHA nodes, and a mix of TDMA nodes and ALOHA nodes, respectively. The optimal results for benchmarking purposes can also be derived by imagining a model-aware node for different cases (the derivations are provided in \cite{benchmark} and omitted here.) 
\subsection{Coexistence with TDMA networks}
We first present the results of the coexistence of one DRL node with one TDMA node. In this trivial case, achieving proportional fairness is the same as maximizing sum throughput. That is, to achieve proportional fairness, the optimal strategy of the DRL node is to transmit in the slots not occupied by the TDMA node and keep silent in the slots occupied by the TDMA node.  Fig. \ref{fig:agent+TDMA2} presents the results when the number of slots assigned to TDMA node is 2, 3, 7 and 8 out of 10 slots within a frame. We can see that the reformulated DLMA protocol can achieve proportional fairness in this case. 
\subsection{Coexistence with ALOHA networks}
We next present the results of the coexistence of one DRL node with one  $ q $-ALOHA node, one FW-ALOHA node, and one EB-ALOHA, respectively. Fig. \ref{fig:agent+qALOHA2} presents the results with different transmission probabilities for the coexistence of one DRL node with one  $ q $-ALOHA node. Fig. \ref{fig:agent+FW-ALOHA2} presents the results with different fixed-window sizes for the coexistence of one DRL node with one FW-ALOHA node. Fig. \ref{fig:agent+EB-ALOHA2} presents the results with different initial window sizes and   $ m=2 $ for the coexistence of one DRL node with one EB-ALOHA node. As shown in these results, the reformulated DLMA protocol can again achieve proportional fairness without knowing the transmission schemes of different ALOHA variants. 
\subsection{Coexistence with a mix of TDMA and ALOHA networks}
We now present the results of a setup where one DRL node coexists with one TDMA node and one  $ q $-ALOHA node simultaneously. We also consider the two cases investigated in Section \ref{mix1}, but the objective now is to achieve proportional fairness among all the nodes. Fig. \ref{fig:agent+TDMA+qALOHA1_2} and Fig. \ref{fig:agent+TDMA+qALOHA2_2} present the results of the two cases. We can see that with the reformulated DLMA protocol, the individual throughputs achieved approximate the optimal individual throughputs achieved by imaging a model aware node.  

\begin{figure}[!t]
	\centering
	\includegraphics[scale=0.5]{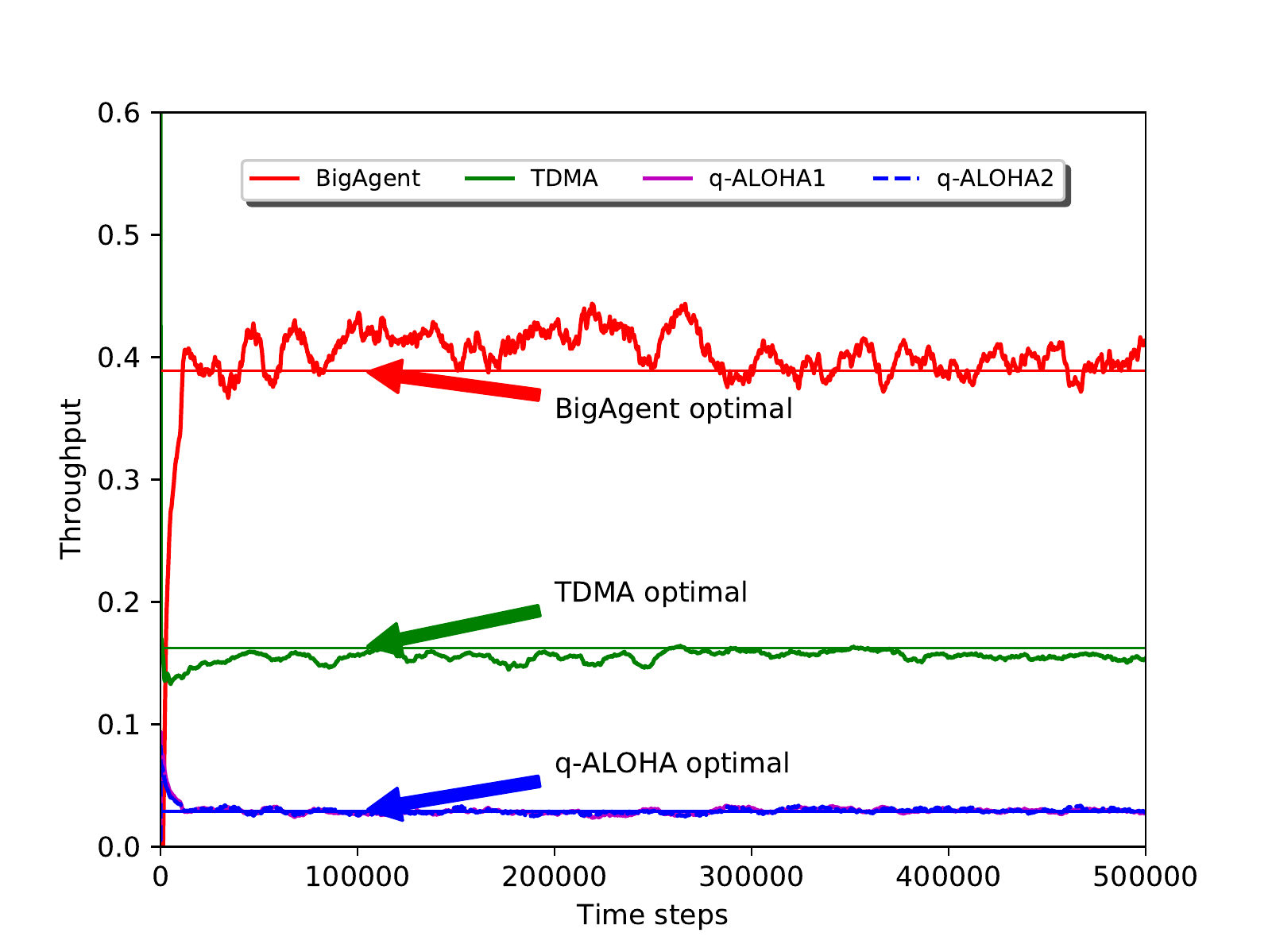}
	\caption{Individual throughputs under the coexistence of three DRL nodes (BigAgent) with one TDMA node and two $  q $-ALOHA nodes. The throughput at time $ t $ is computed  $\sum\nolimits_{\tau  = t - N + 1}^t {{r_\tau }/N} $, where    $ N=5000 $. }
	\label{fig:proportional}
	\vspace{0in}
\end{figure}
We now present the results when three DRL nodes coexist with one TDMA node and two  $ q $-ALOHA nodes. The case investigated here is the same as the case presented in Fig. \ref{fig:sum}, but the three DRL nodes are now formulated to be one big agent and the objective is modified to achieve proportional fairness among all the nodes. The optimal results for the big agent, the TDMA node and each $ q $-ALOHA node are derived in \cite{benchmark}. As shown in Fig. \ref{fig:proportional}, the optimal results can also be approximated using the reformulated DLMA protocol. 

\section{Conclusion}
This paper proposed and investigated a MAC protocol based on DRL for heterogeneous wireless networking, referred to as DLMA. A salient feature of DLMA is that it can learn to achieve an overall objective (e.g., $ \alpha $-fairness objective) by a series of state-action-reward observations while operating in the heterogeneous environment. In particular, it can achieve near-optimal performance with respect to the objective without knowing the detailed operating mechanisms of the other coexisting MACs. 

This paper also demonstrated the advantages of using neural networks in reinforcement learning for wireless networking. Specifically, compared with the traditional RL, DRL can acquire the near-optimal strategy and performance with faster convergence time and higher robustness, two essential properties for practical deployment of the MAC protocol in dynamically changing wireless environments. 

Last but not least, in the course of doing this work, we discovered an approach to generalize the Q-learning framework so that more general objectives can be achieved. In particular, for generality, we argued that we need to separate the Q function and the objective function upon which actions are chosen to optimize. A framework on how to relate the objective function and the Q function in the general set-up was presented in this paper.

\ifCLASSOPTIONcaptionsoff
  \newpage
\fi

\bibliographystyle{IEEEtran}
\bibliography{DLMA}

\end{document}